\documentclass{JHEP3}
\usepackage[normalem]{ulem}
\usepackage{graphicx}
\usepackage{graphics}
\usepackage{cite}
\usepackage{epsfig}
\usepackage{amsmath}
\usepackage{mathrsfs}
\usepackage{amsthm}
\usepackage{amssymb}
\usepackage{oldgerm}
\usepackage{enumerate}
\usepackage{bm,longtable,enumerate}
\usepackage{amsmath,amssymb}
\def\beq{\begin{equation}}
\def\eeq{\end{equation}}
\def\beqa{\begin{eqnarray}}
\def\eeqa{\end{eqnarray}}
\def\ss{\scriptscriptstyle}

\title{Gravitational quasinormal modes of AdS black branes
in $\bf{d}$ spacetime dimensions}

\author{Jaqueline Morgan,$^{a}$ Vitor Cardoso,$^{b,c}$
Alex S. Miranda,$^{d}$ C. Molina,$^{e}$
and \hspace{2cm} Vilson T. Zanchin$^{a}$
\\
$^{a}$Centro de Ci\^encias Naturais e Humanas,
Universidade Federal do ABC,\\
Rua Santa Ad\'elia 166, 09210-170 Santo Andr\'e, SP, Brazil\\
$^{b}$Centro Multidisciplinar de Astrof\'{\i}sica -
CENTRA, Dept. de F\'{\i}sica,\\ Instituto Superior T\'ecnico,\\
Avenida Rovisco
Pais 1, 1049-001 Lisboa, Portugal\\
$^{c}$Department of Physics and Astronomy, The University of
Mississippi, University,\\
MS 38677-1848, U.S.A.\\
$^{d}$Instituto de F\'{i}sica, Universidade Federal do Rio de
Janeiro,\\
Caixa Postal 68528, RJ 21941-972, Brazil\\
$^e$Escola de Artes, Ci\^{e}ncias e Humanidades, Universidade de
S\~{a}o Paulo,\\
Avenida Arlindo Bettio 1000, 03828-000 S\~{a}o Paulo, SP, Brazil\\
E-mail: jaqueline.morgan@ufabc.edu.br, vcardoso@fisica.ist.utl.pt,
astmiranda@if.ufrj.br, cmolina@usp.br, zanchin@ufabc.edu.br}

\abstract{The AdS/CFT duality has established a mapping between quantities
in the bulk AdS black-hole physics and observables in a boundary
finite-temperature field theory. Such a relationship appears to be valid
for an arbitrary number of spacetime dimensions, extrapolating the original
formulations of Maldacena's correspondence. In the same sense properties
like the hydrodynamic behavior of AdS black-hole fluctuations have been
proved to be universal. We investigate in this work the complete
quasinormal spectra of gravitational perturbations of $d$-dimensional
plane-symmetric AdS black holes (black branes). Holographically the
frequencies of the quasinormal modes correspond to the poles of two-point
correlation functions of the field-theory stress-energy tensor. The
important issue of the correct boundary condition to be imposed on the
gauge-invariant perturbation fields at the AdS boundary is studied and
elucidated in a fully $d$-dimensional context. We obtain the dispersion
relations of the first few modes in the low-, intermediate- and
high-wavenumber regimes. The sound-wave (shear-mode) behavior of
scalar (vector)-type low-frequency quasinormal mode is analytically
and numerically confirmed.  These results are found employing both a power
series method and a direct numerical integration scheme.}


\preprint{}

\begin{document}

\section{Introduction}
\label{introd}

The anti-de Sitter/conformal field theory (AdS/CFT)
correspondence \cite{Maldacena:1997re,Witten:1998qj,Gubser:1998bc} has been
widely recognized as an important tool to explore a variety of
non-perturbative aspects of strongly coupled gauge theories. Holographic
string-theory models are now used to study the physics of
strong interactions and condensed matter, both in the zero- and
finite-temperature regimes (see, e.g., Refs.
\cite{BoschiFilho:2006pt,Son:2007vk, Gubser:2007zz,
Erdmenger:2007cm,Iancu:2008sp,Hartnoll:2009sz, Herzog:2009xv} for reviews
and lecture notes on AdS/CFT applications to QCD and
condensed matter physics). One of the
essential ingredients of this approach is the weak/strong relationship
between the coupling constants \cite{Aharony:1999ti}. When the 't Hooft
coupling of the large-$N$ CFT is strong, string theory on AdS spacetime
reduces to classical supergravity theory. In such context one can
investigate diverse phenomena in a class of strongly interacting field
theories by doing the computations on the gravity side of the
correspondence. Among other results, this procedure has allowed
the determination of near-equilibrium properties of the dual CFT plasma,
such as transport coefficients like viscosity, conductivity and
diffusion constants \cite{Policastro:2001yc,
Policastro:2002se,Policastro:2002tn, Herzog:2003ke,Herzog:2002fn,
Kovtun:2004de}.

The AdS/CFT correspondence is also used to
study fundamental questions in gravitational physics,
which are hard or even impossible to be addressed within current gravity
theories, such as the nature of spacetime singularities
\cite{Fidkowski:2003nf,Hubeny:2004cn,Hertog:2004rz,
Hertog:2005hu,Festuccia:2005pi} and the loss of information
in black holes \cite{Hawking:2005kf,Lowe:2006xm}.
Even in regimes for which it is possible to obtain
results from gravity theories, the AdS/CFT correspondence leads to new
interpretations of those results. One example is the evolution of classical
fields in the neighborhood of asymptotically AdS black holes (black
branes). The vectorial sector of gravitational perturbations presents a
fundamental quasinormal mode (QNM) frequency which is purely damped and
goes to zero in the small wavenumber limit. The unusual behavior of this
mode was not understood from a strictly
gravitational point of view
\cite{Cardoso:2001bb,Cardoso:2001vs,Cardoso:2003cj,Miranda:2005qx}.
However, based on the AdS/CFT duality and taking into account the expected
features of low-energy fluctuations in interacting field theories,
Policastro, Son and Starinets \cite{Policastro:2002se,Policastro:2002tn}
were able to interpret such quasinormal mode as the dual of the shear
transverse mode predicted by relativistic fluid mechanics. Some works
\cite{Kovtun:2003wp,Iqbal:2008by} then suggested that the old
`membrane-paradigm' framework
\cite{Thorne:1986iy,Parikh:1997ma,Fujita:2007fg}, in which
the (stretched) horizon is interpreted as a fluid, could be used to explain
the hydrodynamic properties of black holes, and this very important concept
was incorporated in the physics of dynamical classical fields in AdS
spacetimes.

The hydrodynamic behavior of AdS black hole fluctuations, in
particular the universality of such a behavior for four, five,
and seven spacetime dimensions, both at first- and second-order
expansions in the frequency and
momentum \cite{Baier:2007ix,
Natsuume:2007ty,Natsuume:2008iy,Mas:2007ng,kapusta:066017,
springer:086003}, is now a
generally accepted property of AdS black holes (see also Refs.
\cite{Bhattacharyya:2007vs,Bhattacharyya:2008jc,Bhattacharyya:2008xc,
Bhattacharyya:2008mz,Bhattacharyya:2008kq,Bhattacharyya:2008ji,
Fouxon:2008ik,Eling:2009pb} for recent developments in the fluid/gravity
correspondence).

However, there are several other features of the quasinormal spectra of AdS
black holes and black branes which were not considered in higher
dimensional spacetimes, especially in the case of gravitational
perturbations. Such a study is important because, among others, it allows
one to verify if there are aspects of the spectra which are specific to a
given spacetime dimension, or which aspects are dimension-independent. One
example is the crossover from the
hydrodynamic regime to a ``colisionless regime'' appearing in
four and five spacetime dimensions \cite{Nunez:2003eq,
Herzog:2007ij,Amado:2007yr,Miranda:2008vb}.
For any event-horizon size
(or temperature), there is a critical wavenumber value
above which the late-time evolution of the vector-type
gravitational fluctuations is dominated by
the first gapped quasinormal mode and not by the
hydrodynamic shear mode. A possible extension of this crossover for
perturbations of higher-dimensional AdS black holes (and black branes) has
not been investigated yet. This is one of the goals of the present
work.

There are other important issues in the study of the vibrational modes of
AdS black branes to be analyzed in a fully $d$-dimensional
context. We can mention the arbitrariness in the choice of gauge-invariant
combinations of metric variations as fundamental variables of the
gravitational perturbations. Another related issue is the
ambiguity in defining an appropriate condition for the quasinormal modes at
AdS spacetime boundary. Historically, the perturbation variables are
chosen in such a way that the radial part of the fundamental equations
takes a Schr\"odinger-like form when written in terms of the Regge-Wheeler
tortoise coordinate. These are called the Regge-Wheeler-Zerilli
(RWZ) variables. In some of the works on this subject
\cite{Friess:2006kw, Michalogiorgakis:2006jc,Siopsis:2007wn,Alsup:2008fr}
the authors have chosen RWZ type variables and argued that, according to
the AdS/CFT duality, the conditions to be imposed at AdS boundary are such
that gravitational perturbations do not deform the boundary metric. In the
four-dimensional case, Michalogiorgakis and Pufu
\cite{Michalogiorgakis:2006jc} showed that a Robin boundary condition is
the correct condition to be imposed on the RWZ master variable governing
the scalar-type perturbations. With such a boundary condition they were
able to obtain, for instance, the hydrodynamic wave sound mode
which had not been obtained in early works using RWZ variables and a
Dirichlet condition at AdS boundary
\cite{Cardoso:2001bb,Cardoso:2001vs,Cardoso:2003cj,Miranda:2005qx}.
A different route was taken in Refs. \cite{Kovtun:2005ev,
Miranda:2008vb}, where the ambiguities characteristic of
classical-field dynamics at AdS spacetimes were eliminated by defining the
quasinormal (QN) frequencies as the poles, in the space of
frequency and momentum, of retarded Green functions
in the dual field theory. In this approach, the standard tools
to compute real-time Green functions from holography
\cite{Son:2002sd,Herzog:2002pc,Skenderis:2008dh} are used in
order to find the correct boundary conditions that should be imposed
on metric perturbations at the AdS boundary. Any set of perturbation 
functions chosen to fulfill these requirements are called Kovtun-Starinets
(KS) variables. In particular, it was shown that Dirichlet boundary
conditions and KS type variables lead to the correct quasinormal spectra of
AdS black branes in four and five spacetime dimensions
\cite{Kovtun:2005ev,Miranda:2008vb}. These and other related subjects are
investigated here considering AdS black branes in spacetimes of arbitrary
number of dimensions. 

The present work also aims to address other issues. For instance: 
\begin{itemize}
\item[(i)] establish (numerically) the stability of black branes
against
scalar-type perturbations, a result that was proved only for
four-dimensional  black holes and black branes \cite{Kodama:2003ck};
\item[(ii)] analyze the causality of signal propagation in the dual
CFT plasma using
recent results on the eikonal limit of the  QNM spectra
\cite{Festuccia:2008zx,Morgan:2009vg}; 
\item[(iii)] use a time evolution method to
investigate the absence/presence of power-law tails at late stages of the
evolution of perturbations in higher dimensional AdS black branes, and;
\item[(iv)] search for the highly real modes found analytically by
Daghigh and
Green \cite{Daghigh:2009zz,Daghigh:2009fy}, but were not 
confirmed numerically until this moment.
\end{itemize}

The structure of this work is the following. In the next section we define
the $d$-di\-mensional AdS black brane spacetime and the
conventions adopted in the main body of the work. In section
\ref{EQSperturbations} it is presented the one-dimensional
Schr\"odinger-like equations obtained in Ref. \cite{Kodama:2003jz} for
the RWZ master variables. The same section is also devoted to obtain
the fundamental equations for the Kovtun-Starinets variables using the
partially covariant and totally gauge-invariant formalism of Kodama,
Ishibashi and Seto \cite{Kodama:2000fa}.
In section \ref{boundary} we analyze which boundary
conditions should be imposing on KS and RWZ variables
in order to obtain the same spectrum in each perturbation sector. The
analysis is performed for an arbitrary number of spacetime dimensions.
Section \ref{qnmresults} is devoted to report a few interesting analytical
results. The numerical results  are presented and analyzed in section
\ref{secNumericalQNM}. In section \ref{subsecAdSCFT} the QNM
are analyzed in terms of the AdS/CFT correspondence, and in the
section \ref{secfinal} we make final comments and conclude.

\section{The background spacetime}\label{secBack}

The background spacetime considered here represents a $d$-dimensional
plane-symmetric asymptotically anti-de Sitter (AdS) black hole, or
simply an AdS black brane
\cite{Lemos:1994fn,Huang:1995zb,Lemos:1994xp,Cai:1996eg,Lemos:1995cm,
Awad:2002cz}. The spacetime can be locally written as
a product of a two-dimensional spacetime ${\cal N}^2$, spanned
by a timelike coordinate $t$ and a radial spacelike
coordinate $r$, and a $(d-2)$-dimensional space ${\cal K}^{d-2}$
with constant sectional curvature $K=0$ \cite{Kodama:2000fa,Kodama:2003jz}. 
With such a decomposition, the background metric in Schwarzschild-like
coordinates takes the form
\begin{equation}\label{fundo}
ds^2\;=\;\frac{r^{2}}{R^{2}}\;\left[-f(r)\;dt^{2}
+\sum^{d-1}_{i=2}\;dx^{i}dx_{i}\;\right]+\frac{R^{2}}{r^{2}
f(r)}\;dr^{2},
\end{equation}
for which
\begin{equation}\label{76}
f(r)=1-\frac{r_{h}^{d-1}}{r^{d-1}},
\end{equation}
with $r_{h}$ being the event horizon radius, and $R$ the AdS
radius. The coordinates $x^i$, $i=2,3,..., d-1,$
span the ${\cal K}^{d-2}$ space.

The Hawking temperature of the black brane is
 \begin{equation}\label{hawkingTemp}
T=\frac{(d-1)r_{h}}{4 \pi R^{2}},
\end{equation}
and the AdS radius $R$ is given in terms of the negative cosmological
constant $\Lambda$ through the relation
 \begin{equation}\label{deSitterradius}
R^{2}=-\frac{(d-2)(d-1)}{2 \Lambda}.
\end{equation}

The radial coordinate $r$ covers, without singularities, the whole region
of interest for the analysis of the QNM of the AdS black hole of metric
\eqref{fundo}, namely, the range $(r_h,\infty)$. To simplify the analysis,
as usual we introduce a new coordinate which is defined in a finite
interval. This is done through the following re-parameterization
\begin{equation}\label{u_def}
u=\frac{r_{h}}{r},
\end{equation}
which results in
\begin{equation}\label{f_u}
f(u)=1-u^{d-1}\, .
\end{equation}
Now the event horizon is located at $u=1$, and the AdS spatial infinity
($r \rightarrow\infty$) is at $u=0$. Hence, we have $u\in (1,0)$,
and the metric \eqref{fundo} becomes
\begin{equation}\label{fundo2}
ds^2\;=\;\frac{r_{h}^2}{u^{2}R^{2}}\;\left[-f(u)\;dt^{2}
+\sum^{d-1}_{i=2}\;dx^{i}dx_{i}\;\right]+\frac{R^{2}}{u^2f(u)}\; du^{2}.
\end{equation}
In the following, coordinates $(t,u)$ are labeled as $x^a$, $a=0,\,1$,
i.e., coordinates $t$ and $u$ span the relevant region of ${\cal N}^2$
outside the horizon.

The foregoing black-brane spacetime has been extensively studied
in the last years in connection with the AdS/CFT correspondence,
specially for $d=4,$ $5$ and $7$ dimensions. In such a cases
the metric \eqref{fundo2} can be seen as part of nonextremal solutions
to the supergravity equations of motion in ten or eleven dimensions
\cite{Aharony:1999ti,Horowitz:1991cd}. The near-horizon
limit of the full supergravity spacetime is the direct product
of an $\mbox{AdS}_{d}$ black brane and a $S^{D-d}$ sphere,
where $D=10$ for $d=5$ and $D=11$ for $d=4$ and $7$. The internal
degrees of freedom corresponding to the $(D-d)$-dimensional sphere
will not be important for the present work, since we are interested
in the correlators of the CFT energy-momentum tensor and, according
to the gauge/gravity dictionary \cite{Son:2007vk,Berti:2009kk},
this operator is dual to the gravitational fluctuations of the
background spacetime \eqref{fundo2}.

The general properties of metric perturbations of the
considered $d$-dimensional black branes are investigated in the next
section where we write the fundamental equations that govern the
evolution of gravitational perturbations in these asymptotically AdS
spacetimes.

\section{Fundamental equations for the gravitational
perturbations} \label{EQSperturbations}

Following the procedure  presented in Ref. \cite{Kodama:2000fa}, the
gravitational perturbations are expanded in terms of harmonic functions
on ${\cal K}^{d-2}$ and the first-order perturbed Einstein equations
are given in terms of a set of gauge-invariant quantities. These quantities
are combinations of the metric perturbations $h_{\alpha\beta}$ which
are related to the perturbed spacetime metric through the usual
definition $g_{\alpha\beta}= g^{\mbox{\tiny{$(0)$}}}_{\alpha\beta}
+h_{\alpha\beta}$, where $g^{\mbox{\tiny{$(0)$}}}_{\alpha\beta}$
stands for the background metric defined by Eq. \eqref{fundo2}. The
gravitational perturbations are grouped into three distinct classes
(sectors) according to the special type of harmonic tensors
that appear in the expansions of $h_{\alpha\beta}$. These can be tensorial,
vectorial, or scalar perturbations, corresponding respectively to the
scalar, shear and sound symmetry channels for the gravitational
fluctuations considered in Ref. \cite{Kovtun:2005ev}.
Each one of these sectors is governed by a particular closed
group of independent differential equations. It is possible to choose a
particular set of master variables which allows to write only one
perturbation equation for each perturbation sector, as, for instance, the
RWZ set of variables adopted in Ref. \cite{Kodama:2003jz}. Another
interesting set of variables (KS variables) were first used in Ref.
\cite{Kovtun:2005ev}. Here we present the fundamental equations
for KS variables in $d$ spacetime dimensions, and establish
a connection between RWZ and KS variables.
The gauge-invariant metric perturbation nomenclature and labelling follows
Ref. \cite{Kodama:2000fa}.

\subsection{Metric perturbations}

\subsubsection{Tensorial sector}
\label{subsectpertTen}

This particular set of gravitational perturbations can be
represented in terms of tensorial harmonics
$\mathbb{T}_{ij}$, in the form
\begin{equation}\label{tensorial-pert}
h_{ab}=0,\quad h_{ai}=0, \quad
h_{ij}=2u^{-2}H_{\ss{T}}\mathbb{T}_{ij},
\end{equation}
where $H_{\ss{T}}=H_{\ss{T}}(t,u)$ is a gauge-invariant
function depending on the coordinates $t$ and $u$ only, and
$\mathbb{T}_{ij}$ are transverse traceless
harmonic tensors defined on ${\cal K}^{d-2}$ \cite{Kodama:2000fa}.

\subsubsection{Vectorial sector}
\label{subsectpertVec}

Metric perturbations of vectorial type can be expanded in terms of
vectorial harmonic functions $\mathbb{V}_{i}$ as follows
\begin{equation}\label{vectorial-pert}
h_{ab}=0,\quad h_{ai}=u^{-1}f_{a}\mathbb{V}_{i},
\quad h_{ij}=2u^{-2}H_{\ss{V}}\mathbb{V}_{ij},
\end{equation}
where $f_a=f_{a}(t,u)$ and $H_{\ss{V}}= H_{\ss{V}}(t,u)$ are scalar
functions of the coordinates on ${\cal N}^2$, to be determined, and
$\mathbb{V}_{ij}$ are vector-type harmonic tensors on ${\cal K}^{d-2}$
built from the transverse harmonic vectors $\mathbb{V}_{i}$ (see Ref.
\cite{Kodama:2000fa}). From the functions $f_{a}$ and $H_{\ss{V}}$ it
is defined a new set of gauge-invariant quantities $F_{a}$
($a=0,\,1$) given by
\begin{equation}\label{ginvar-vector}
F_{a}=f_{a}+\frac{1}{uk}D_{a}H_{\ss{V}}\, ,
\end{equation}
where $k$ is the perturbation wavenumber, and $D_a$ is the
covariant derivative in the space ${\cal N}^2$.

\subsubsection{Scalar sector}\label{subsecpertScal}

Scalar gravitational perturbations are the set of metric perturbations
which can be expanded in terms of scalar harmonic functions $\mathbb{S}$
in the form  
\begin{equation}\label{scalar-pert}
h_{ab}=\mathfrak{f}_{ab}\,\mathbb{S},\qquad
h_{ai}=u^{-1}\mathfrak{f}_{a}\,\mathbb{S}_{i},\qquad h_{ij}
=2u^{-2}\left(H_{\ss{L}}\gamma_{ij}\mathbb{S}+H_{\ss{S}}
\mathbb{S}_{ij}\right),
\end{equation}
where $\mathfrak{f}_{ab}=\mathfrak{f}_{ab}(t,u)$,
$\mathfrak{f}_{a}=\mathfrak{f}_{a}(t,u)$, $H_{\ss{L}}=H_{\ss{L}}(t,u)$ and
$H_{\ss{S}}=H_{\ss{S}}(t,u)$ are functions to be determined.
$\mathbb{S}_{i}$ and $\mathbb{S}_{ij}$ are respectively scalar-type
harmonic vectors and tensors on ${\cal K}^{d-2}$ built from the scalar
harmonic functions $\mathbb{S}$ (see Ref. \cite{Kodama:2000fa}). A set of
gauge-invariant quantities are then defined as
\begin{equation}\label{ginvar-scalar}
\begin{split}
F&=H_{\ss{L}}+\frac{1}{n}H_{\ss{S}}+uD^{a}\left(\frac{1}{u}\right)X_{a},\\
F_{ab}&=\mathfrak{f}_{ab}+D_{a}X_{b}+D_{b}X_{a},
\end{split}
\end{equation}
with
\begin{equation}\label{ginv-scalar2}
X_{a}=\frac{1}{uk}\left(\mathfrak{f}_{a}+\frac{1}{uk}D_{a}H_{\ss{S}}\right).
\end{equation}

Now we write the gravitational fundamental equations for each
perturbation sector and each set of variables, RWZ and KS.

\subsection{Master equations for the RWZ variables}

Kodama and Ishibashi \cite{Kodama:2003jz} showed that for a black brane in
four or more spacetime dimensions, the Einstein equations for
the gravitational perturbations can be reduced to three independent
second-order wave equations in a two-dimensional static spacetime, one
equation corresponding to each one of the perturbation modes. Moreover, the
variable for the final second-order master equation for a specific
mode is given by a simple combination of gauge-invariant quantities in the
formalism of Ref. \cite{Kodama:2000fa}. It is introduced new variables
$\Phi_{p}$ so that, after Fourier decomposition of such perturbation
functions, $\Phi_{p}(t,u)=\int{\Phi_{p}(u)\, e^{i\omega t}d\omega}$, the
perturbation equations take a Schr\"odinger-like form, 
\begin{equation}\label{fund-eq-RWZ}
\frac{d^{2}\Phi_{p}}{d\mathfrak{r}_{*}^{2}}+(\mathfrak{w}^{2}-V_{p})
\Phi_{p}=0,
\end{equation}
where $\mathfrak{r}_{*}$ is the normalized tortoise radial coordinate,
defined by $du/d\mathfrak{r}_{\ast}=-f(u)$. The label $p$
can be {\small{\it{T, V}}} or {\small{\it{S}}} depending of the
perturbation sector: tensorial, vectorial and
scalar, respectively.
$V_{p}$ is the effective potential, and the parameter $\mathfrak{w}$
is the normalized frequency defined by  
\begin{equation}\label{normfreq}
 \mathfrak{w}=\frac{(d-1)\,\omega}{4 \pi T}=\dfrac{R^2}{r_h}\, \omega ,
\end{equation}
where $T$ stands for the Hawking temperature of the black brane.

Next we define the RWZ variable $\Phi_p$ for each perturbation
sector and the corresponding effective potentials.

\subsubsection{Tensorial sector}

As argued in Ref. \cite{Kodama:2003jz}, the simplest function 
$\Phi_{\ss{T}}$ that allows to write the resulting perturbation equation 
in a Schr\"odinger-like form is
\begin{equation}
 \Phi_{\ss{T}}= u^{-\frac{d-2}{2}} H_{\ss{T}}\,,
\end{equation}
where $H_{\ss{T}}$ (introduced in Eqs.~\eqref{tensorial-pert}) is a
gauge-invariant quantity by itself \cite{Kodama:2000fa}. In such a case,
the potential $V_{\ss{T}}$ (cf. Eq.~\eqref{fund-eq-RWZ}) for this sector
is given by
\begin{equation}\label{pott}
V_{\ss{T}}(u)=f(u)\left[\mathfrak{q}^{2}+\frac{d(d-2)}{4u^{2}}
+\frac{(d-2)^{2}u^{d-3 } } {4 } \right].
\end{equation}
Here the parameter $\mathfrak{q}$ is the normalized wavenumber defined by  
\begin{equation}\label{normwavenum}
 \mathfrak{q}=\frac{(d-1)\,k}{4 \pi T}=\dfrac{R^2}{r_h}\, k .
\end{equation}

\subsubsection{Vectorial sector}

The variable $\Phi_{\ss{V}}$ is defined implicitly by (see
Ref. \cite{Kodama:2003jz})
\begin{equation}
 F^{a}=u^{d-3}\,\epsilon^{ab}D_{b}\left(u^{-\frac{d-2}{2}}
\Phi_{\ss{V}}\right),
\end{equation}
where $F^{a}$ is the gauge-invariant quantity defined
in Eq. \eqref{ginvar-vector} \cite{Kodama:2000fa}, and
$\epsilon^{ab}$ is the Levi-Civita tensor in the two-space 
${\cal N}^{\,2}$.
The corresponding effective potential for the vectorial sector
$V_{\ss{V}}$ is
\begin{equation}\label{potv}
V_{\ss{V}}(u)=f(u)\left[\mathfrak{q}^{2}+\frac{(d-2)(d-4)}{4u^{2}}-
\frac{3(d-2)^{2}u^{d-3}}{4}\right].
\end{equation}

\subsubsection{Scalar sector}

In this sector, the RWZ variable $\Phi_{\ss{S}}$ suggested in Ref.
\cite{Kodama:2003jz} is given by
\begin{equation}
\Phi_{\ss{S}}=\frac{2(d-2)\,u^{-\frac{d-4}{2}}}
 {2\mathfrak{q}^{2}+{(d-1)(d-2)u^{d-3}}}
\left(\frac{2F}{u}+\frac{if(u)F_{ut}}{\mathfrak{w}}\right),
\end{equation}
where $F$ and $F_{ut}$ are the gauge-invariant quantities given by
Eqs.~\eqref{ginvar-scalar} \cite{Kodama:2000fa}, and the effective
potential is

\begin{equation}\label{pots}
 V_{\ss{S}}(u)=\frac{f(u)Q(u)}{4\left[2\mathfrak{q}^{2}+{(d-1)(d-2)u^{d-3}}
\right]^2}.
\end{equation}
Here $Q(u)$ is given by
\begin{equation}
\begin{split}
 Q(u)=&(d-2)^{3}\left[d+(d-2)(1-f)\,\right] \frac{{f'}^2}{u^4}-
4(d-2)\left[(d-5)(d-2)(d-1)+(d-2)^2 f\right.\\
&\left.-4(1-f)\right]\frac{f'}{u^3}\mathfrak{q}^{2 }
 +4(d-6)\left[d-4-3(d-2)(1-f)\right]\frac{1}{u^2}\mathfrak{q}^{4}
+16\mathfrak{q}^{6}.
\end{split}
\end{equation}

\subsection{Master equations for the KS variables}

Another choice of fundamental variables for the gravitational
perturbations was firstly suggested by Kovtun and Starinets
\cite{Kovtun:2005ev}. A set of master equations for the Kovtun-Starinets
(KS) variables in $d=4$ and $5$ dimensions were obtained
in Refs. \cite{Miranda:2008vb,Kovtun:2005ev}. In connection with
the formalism of Ref. \cite{Kodama:2000fa}, we present here the fundamental
equations for the KS variables in $d$ spacetime dimensions.

\subsubsection{Tensorial sector}

For $d$ spacetime dimensions the KS variable for the tensorial sector
$Z_{\ss{T}}$  is defined by
\begin{equation}
 Z_{\ss{T}}=H_{\ss{T}}/2,
\end{equation}
where $H_{\ss{T}}$ is the gauge-invariant quantity introduced in
Eqs.~\eqref{tensorial-pert}. In terms of $Z_{\ss{T}}(u)$, defined by
$Z_{\ss{T}}(t,u)=\int{Z_{\ss{T}}(u)\, e^{i\omega t}d\omega}$,
the perturbation equation for the tensorial sector
is given by
\begin{equation}\label{fund-eq-tensor}
Z_{\ss{T}}''-\left[\frac{d-1 -f }{u f}\right]Z_{\ss{T}}'
+\left[\frac{\mathfrak{w}^2
-\mathfrak{q}^{2}f}{f^2}\right]Z_{\ss{T}}=0,
\end{equation}
where the primes indicate derivatives with respect to the coordinate $u$,
and $f=f(u)$ is the horizon function defined in Eq. \eqref{f_u}.
This equation reduces to the corresponding perturbation equation of Ref. 
\cite{Kovtun:2005ev} when one takes $d=5$ and makes the
adjustments for different notation and normalizations.

\subsubsection{Vectorial sector}

In connection with the formalism developed by Kodama, Ishibashi and
Seto \cite{Kodama:2000fa}, the KS master variable for the vectorial
gravitational perturbations takes the form
\begin{equation}
 Z_{\ss{V}}={F}_{t}/u,
\end{equation} 
where $F_{t}$ is the gauge-invariant quantity of Ref.
\cite{Kodama:2000fa}, as defined in Eq.~\eqref{ginvar-vector}. With
this variable, after Fourier decomposition as in the tensorial case,
we obtain the following equation
\begin{equation}\label{fund-eq-vector}
Z_{\ss{V}}''-\left[\frac{d-2}{u} +\frac{f'\mathfrak{w}^{2}}
{f(\mathfrak{q}^2f-\mathfrak{w}^{2})}\right]Z_{\ss{V}}'+
\left[\frac{\mathfrak{w}^2-\mathfrak{q}^{2}f}{f^2}\right]Z_{\ss{V}}=0,
\end{equation}
which is the master equation for the vectorial metric perturbations of
$d$-dimensional black branes in terms of the KS gauge-invariant
variable $Z_{\ss{V}}$. The general expression Eq. \eqref{fund-eq-vector}
reduces to the equations for vector perturbations in four and five spacetime
dimensions, as seen in \cite{Miranda:2008vb, Kovtun:2005ev}.

\subsubsection{Scalar sector}

Again inspired in the work by Kovtun and Starinets
\cite{Kovtun:2005ev}, we write the following
gauge-invariant quantity to describe the scalar-type gravitational
perturbations,
\begin{equation}\label{ZFF}
Z_{\ss{S}}= u^2 F_{tt}+ \left[(d-1)u^{d-1} +2f(u)\right]F\,,
\end{equation}
where $F_{tt}$ and $F$ are gauge-invariant quantities defined in
Eqs.~\eqref{ginvar-scalar} \cite{Kodama:2000fa}. With this
expression it is shown that $Z_{\ss{S}}$ satisfies the
following differential equation
\begin{equation}\label{fund-eq-scalar}
Z_{\ss{S}}''+ \frac{Y_1\mathfrak{q}^{2}+Y_2\mathfrak{w}^{2}}
{uf\, X}\,Z_{\ss{S}}'+\frac{Y_3\mathfrak{q}^{2}+ 
Y_4\mathfrak{q}^{4}+2\,(d-2)\,\mathfrak{w}^{4}}{f^{2}\,X}\, Z_{\ss{S}}=0,
\end{equation}
where we have introduced the coefficients
\begin{equation*}
\begin{split}
X\, &=2\,(d-2)\,\mathfrak{w}^{2}-\left[d-3+(d-1)f\right]\mathfrak{q}^{2},
\\
Y_1 &=2(d-2)^{2}f^2+(d-1)\left(d-1+f\right)u^{d-1}, \\
Y_2 &=-2(d-2)\left[d-1-f\right],\\
Y_3 &=-(d-3){f'}^2f -\left[4(d-2)f+(d-1)u^{d-1}\right] \mathfrak{w}^{2},\\
Y_4 &=\left[2(d-2) + (d-1)u^{d-1}\right]f.
\end{split}
\end{equation*}
It is important to mention that Eq. \eqref{fund-eq-scalar}
can be reduced to the scalar perturbation equation of
\cite{Miranda:2008vb} when one takes $d=4$, and to the corresponding
equation of \cite{Kovtun:2005ev} when one takes $d=5$.

\section{Gauge-invariant variables and boundary conditions}
\label{boundary}

It is known that ingoing wave condition at horizon and
Dirichlet condition at AdS boundary applied to the
Regge-Wheeler-Zerilli (RWZ) variables does not give all the QNM
of a given perturbative sector due to the choice
of the boundary condition at spatial infinity \cite{Miranda:2008vb}.
In particular, the sound wave mode in
four dimensional spacetimes does not show up
\cite{Cardoso:2001bb,Cardoso:2001vs,Cardoso:2003cj,Miranda:2005qx}.
On the other hand, using the same boundary conditions and Kovtun-Starinets
(KS) gauge-invariant variables the mentioned sound wave mode appeared
\cite{Herzog:2003ke,Miranda:2008vb}, and it was also verified that
in some cases RWZ and KS yield different non-hydrodynamic
quasinormal frequencies \cite{Miranda:2008vb}. In these circumstances
one must be able to decide which spectrum has a meaningful physical
interpretation. In accordance to the AdS/CFT correspondence we opt for
the QN frequencies obtained from the poles of the related two-point
correlation functions, i.e., we choose the spectrum obtained by
applying an ingoing wave condition at horizon and a
Dirichlet boundary condition at $u=0$ to
the KS gauge-invariant variables $Z_p(u)$ \cite{Kovtun:2005ev}. 
 However, it is known that different
master variables can lead in special cases to the same spectrum,
as it happens with the polar and axial gravitational
perturbations of asymptotically flat four-dimensional black holes
\cite{MTB}. Having this in mind the objective in this section
is to investigate which boundary conditions must be applied to each
variable in order to produce the QNM spectrum corresponding to
the poles of the stress-energy tensor correlators
in the dual field theory. A comparison among the
spectra obtained with the same boundary conditions applied to the RWZ
quantities $\Phi_p(u)$ and to the KS variables $Z_p (u)$ is also done. 
The first step is then to find the asymptotic form of the perturbation
functions $Z_{p}$, i.e., we try solutions of the form $Z_{p} \sim u^\nu$,
where $\nu$ is a parameter to be determined. We find that
Eqs.~\eqref{fund-eq-tensor}, \eqref{fund-eq-vector} and
\eqref{fund-eq-scalar} are satisfied in the limit $u\rightarrow0$ if
$\nu=0$, or if $\nu=d-1$. Therefore the solutions for $Z_{p}(u)$
which satisfy the incoming-wave condition at horizon, here denoted by
$Z^{\ss{in}}_{p}(u)$, present the following asymptotic behavior around
$u=0$:
\begin{equation}
Z^{\ss{in}}_{p}(u)=\mathcal{A}_{p}(\mathfrak{w},\mathfrak{q})
+...\;+\mathcal{B}_{p}(\mathfrak{w},\mathfrak{q})u^{d-1}
+...\, ,    \label{asymptoticZ}
\end{equation}
where $p=${\small{\it{T, V, S}}} refers respectively to the tensorial,
vectorial, and scalar perturbation sectors. The ellipses in the
foregoing equation denote higher powers of $u$, and quantities
$\mathcal{A}_{p}(\mathfrak{w},\mathfrak{q})$
and $\mathcal{B}_{p}(\mathfrak{w},\mathfrak{q})$ are the connection
coefficients related to the respective differential equations. After
Eqs.~\eqref{asymptoticZ} one finds that Dirichlet boundary conditions
imposed on $Z^{\ss{in}}_{p}(u)$ give
\begin{equation}
 Z^{\ss{in}}_{p}(0)=\mathcal{A}_{p}(\mathfrak{w},\mathfrak{q}) =0,
\qquad\quad  p=\mbox{{\small{\it{T, V, S}}}}.  \label{spectraZ}
\end{equation}

The next step is to study the relations among the KS $Z_{p}(u)$
and the RWZ $\Phi_{p}(u)$ variables at the AdS boundary and to find the
relations between the QNM spectra, which we call 
the KS- and the RWZ-spectra, respectively, for short. We do that by
considering separately each one of the perturbation sectors.

\subsection{Tensorial sector}\label{varten}

For the tensorial gravitational sector, we were able to find an explicit
relation between the RWZ variable $\Phi_{\ss{T}}(u)$ and the KS
gauge-invariant variable $Z_{\ss{T}}(u)$ for any spacetime dimension. 
It is given by
\begin{equation}
Z_{\ss{T}}(u)=\frac{1}{2}\,u^{\frac{d-2}{2}}\Phi_{\ss{T}}(u)\, . \label{Z3Phi3}
\end{equation}
Furthermore, it can be shown from Eqs.~\eqref{fund-eq-RWZ}
and \eqref{pott} that at the asymptotic region $\Phi^{\ss{in}}_{\ss{T}}(u)$
is of the form
\begin{equation} \label{asymptPhiT}
 \Phi^{\ss{in}}_{\ss{T}}(u)=\mathcal{C}_{\ss{T}}(\mathfrak{w},
\mathfrak{q})u^{-\frac{d-2}{2}}
+...\;+\mathcal{D}_{\ss{T}}(\mathfrak{w},\mathfrak{q})u^{\frac{d}{2}}
+...,
\end{equation}
agreeing with the asymptotic form for $d$ spacetime dimensions
found in
Refs.~\cite{Friess:2006kw,Michalogiorgakis:2006jc}. The asymptotic expressions
for $\Phi_{\ss{V}}$ and $\Phi_{\ss{S}}$ obtained below
(see Eqs. \eqref{Phin}, \eqref{Phi-2ddasympt}, and \eqref{Phi-25dasympt})
are also in accordance with those found in
Refs.~\cite{Friess:2006kw,Michalogiorgakis:2006jc}.
As mentioned above, the Dirichlet boundary condition at
$u=0$ imposed on $Z^{\ss{in}}_{\ss{T}}(u)$ requires that
${\mathcal{A}_{\ss{T}}}(\mathfrak{w},\mathfrak{q})=0$, which is the same as
the condition one obtains by imposing Dirichlet boundary condition on the
RWZ variable $\Phi^{\ss{in}}_{\ss{T}}(u)$. Namely, the relations
\eqref{asymptoticZ}, \eqref{Z3Phi3} and \eqref{asymptPhiT} imply in
${\mathcal{A}_{\ss{T}}}(\mathfrak{w},\mathfrak{q})
=\mathcal{C}_{\ss{T}}(\mathfrak{w}, \mathfrak{q})=0$. Now since the
equation ${\mathcal{A}_{\ss{T}}}(\mathfrak{w},\mathfrak{q})=0$ furnishes
the spectrum of the QNM one concludes that the spectra of the tensorial
gravitational QNM obtained by using KS or RWZ variables
are identical for all dimensions $d>4$.

\subsection{Vectorial sector}\label{varvet}

In the case of gravitational vectorial perturbations, we can
show that the KS and RWZ variables are related by
\begin{equation}
Z_{\ss{V}}(u)=f u^{d-2} \frac{\partial}{\partial
u}\left(u^{-\frac{d-2}{2}}\Phi_{\ss{V}}(u)\right).  \label{Z1Phi1}
\end{equation}
Now Eqs.~\eqref{fund-eq-RWZ} and \eqref{potv} yield the following
asymptotic form for the solution $\Phi^{\ss{in}}_{\ss{V}}(u)$ which
satisfy an incoming-wave condition at the horizon:
\begin{equation}  
\Phi^{\ss{in}}_{\ss{V}}(u)={\mathcal{C}}_{\ss{V}}(\mathfrak{w},
\mathfrak{q})u^{-\frac{
d-4}{2}}+...+ { \mathcal { D } }
_{\ss{V}}(\mathfrak{w},\mathfrak{q})u^{\frac{d-2}{2}}+...\,. \label{Phin} 
\end{equation}
\noindent
Therefore, Eqs.~\eqref{asymptoticZ}, \eqref{Z1Phi1} and \eqref{Phin},
and the Dirichlet boundary condition at $u=0$ imposed on
$Z^{\ss{in}}_{\ss{V}}(u)$ imply in
${\mathcal{A}_{\ss{V}}}(\mathfrak{w},\mathfrak{q})=-(d-3)
{\mathcal{C}_{\ss{ V}}} (\mathfrak{w},\mathfrak{q})= 0$, from what one
concludes that the KS and the RWZ quasinormal spectra are identical to each
other for all dimensions $d\geq 4$.

\subsection{Scalar sector}

For the scalar gravitational sector we were not able to find a simple
relation between $Z_{\ss{S}}(u)$ and $\Phi_{\ss{S}}(u)$, and then the
analysis becomes more evolved than for the other sectors. After some
algebra we find 
\begin{equation}\label{Z2Phi2n}
\begin{split}
&2(d-2)\left[2\mathfrak{q}^{2}u-(d-2)\,f'\right]^2 Z_{\ss{S}}(u)=
\left\{-2(d-2)^{3}{f^\prime}^2\,\mathfrak{w}^{2}\right.\\
&\left.+(d-2)\left[(d-2)\,g(u)\,
u^{d-3}+4\,h(u)\,\mathfrak{q}^2\right]\mathfrak{q}^2 +
4\left[2(d-2)-(d-3)u^{d-1}\right]u^{2}\,\mathfrak{q}^{6}\,\right\} 
u^{\frac{d-2}{2}} \Phi_{\ss{S}}(u)\\
&+\left(2\mathfrak{q}^{2}u-(d-2)\,f'\right)
\left[(d-1)(1-f) -4(d-2)(d-3)f^2 \right]\,
\mathfrak{q}^{2}\,u^{\frac{d-2}{2}}\Phi_{\ss{S}}'(u),
\end{split}
\end{equation}
where the coefficients $h(u)$ and $g(u)$ are defined by
\begin{align}
h(u)=&\,2+(d-2)(d-5)-\left[3+(d-10)(d-2)\right]u^{d-1}\notag\\
&+(d-4)(d-3)u^{2(d-1)}-2u^{2}
\mathfrak{w}^{2},\\
g(u)=&\,2\left[1+(d-2)^{2}+2(d-2)^3 \right]
 u^{d-1}-(d-3)^{2}(d-1)u^{2(d-1)}\notag\\
&-2(d-1)\left[(d-2)(d-3)+ 4u^{2}\mathfrak{w}^{2}\right].
\end{align}

Using Eqs. \eqref{fund-eq-RWZ} and \eqref{pots} for the RWZ scalar variable
$\Phi_{\ss{S}}(u)$ we find the following asymptotic form for the incoming-wave
solution at the horizon $\Phi^{\ss{in}}_{\ss{S}}$:
\begin{align}
 & \Phi^{\ss{in}}_{\ss{S}}(u)={\mathcal{C}_{\ss{S}}}(\mathfrak{w},
\mathfrak{q})u^{-\frac{d-6}{2}}+...
+{\mathcal{D}_{\ss{S}}}(\mathfrak{w},\mathfrak{q})u^{\frac{d-4}{2}}+...,
\qquad d\neq 5, \label{Phi-2ddasympt}\\
&\Phi^{\ss{in}}_{\ss{S}}(u)=\left[{\mathcal{C}}_{\ss{S}}(\mathfrak{w},
\mathfrak { q } )+...+{\mathcal{D}}_{\ss{S}}(\mathfrak{w},\mathfrak{q})\;
\mbox{ln} u+...\right]\sqrt{u} , \quad\qquad d=5,
\label{Phi-25dasympt}
\end{align}
where $\mathcal{C}_{\ss{S}}$ and $\mathcal{D}_{\ss{S}}$ are the
connection coefficients associated to Eq. \eqref{fund-eq-RWZ}.

Since the asymptotic forms of $Z^{\ss{in}}_{\ss{S}}$ and
$\Phi^{\ss{in}}_{\ss{S}}$
critically depend on the number of dimensions we analyze the cases $d=4$,
$d=5$ and $d>5$ separately.

\subsubsection{Four dimensions}

In the four-dimensional case ($d=4$), Eqs.~\eqref{Z2Phi2n} and
\eqref{Phi-2ddasympt}, and the Dirichlet boundary condition on
$Z^{\ss{in}}_{\ss{S}}(u)$ lead to the condition
\begin{equation} \label{RobinC}
 {\mathcal{C}}_{\ss{S}}(\mathfrak{w},\mathfrak{q})
+\frac{3{\mathcal{D}}_{\ss{S}}(\mathfrak{w},\mathfrak{q})}
{\mathfrak{q}^{2}}=0.
\end{equation}
In terms of the RWZ variable $\Phi^{\ss{in}}_{\ss{S}}(u)$, this is a
boundary condition of Robin type, i.e., a mixing between Dirichlet and
Neumann boundary conditions. Therefore, in order for both of the spectra
being the same, and in order for the QN frequencies being given by the
poles of the dual stress-energy tensor correlator, one must impose
Dirichlet boundary condition on $Z^{\ss{in}}_{\ss{S}}(u)$ at $u=0$
and Robin boundary condition on $\Phi^{\ss{in}}_{\ss{S}}(u)$ at $u=0$.
This result explains why Dirichlet boundary conditions imposed on
$Z^{\ss{in}}_{\ss{S}}(u)$ and $\Phi^{\ss{in}}_{\ss{S}}(u)$
lead to different quasinormal spectra. In
Ref.~\cite{Michalogiorgakis:2006jc} it was argued that the
non-deformation of the boundary metric favors a Robin condition on the master
field $\Phi_{\ss{S}}(u)$, and using such a boundary condition
they have found the hydrodynamic QNM of the scalar gravitational sector
in the $d=4$ Schwarzschild-AdS spacetime. Our result is consistent with
that analysis, since both of the results are identical for large
$r_{h}/R$, a regime where the Schwarzschild-AdS black hole
reduces to the AdS black brane. As a matter of fact, it can be shown that
for a spacetime in which the subspace ${\cal K}^{2}$ has constant
curvature $K$ and the event horizon is such that $r_{h}\gg R$,
the relation \eqref{RobinC} is replaced by
$\mathcal{C}_{\ss{S}}
+3\mathcal{D}_{\ss{S}}/(\mathfrak{q}^2-2K{R^2}/{r_h^2})=0$,
which reproduces our result for $K=0$, and the result of
Ref.~\cite{Michalogiorgakis:2006jc} for $K=1$ and
$\mathfrak{q}^2=l(l+1)R^2/r_h^2$, where $l$ is the angular momentum
of the perturbation.

\subsubsection{Five dimensions}
\label{fivedimensions}

The asymptotic form for $\Phi^{\ss{in}}_{\ss{S}}(u)$ for $d=5$ is given
by Eq.~\eqref{Phi-25dasympt}. The Dirichlet boundary condition on 
$Z^{\ss{in}}_{\ss{S}}(u)$ at $u=0$, together with Eqs.~\eqref{Z2Phi2n}
and \eqref{Phi-25dasympt}, furnishes the following condition
\begin{equation} \label{BC-KS-scalar}
 {\mathcal{D}}_{\ss{S}}(\mathfrak{w},\mathfrak{q})=0.
\end{equation}
This is equivalent to the condition of
not changing the metric on the AdS boundary, as shown in Ref.
\cite{Friess:2006kw} in the case of a spherically symmetric $\mbox{AdS}_{5}$
black hole. Hence, using Eqs.~\eqref{BC-KS-scalar} and \eqref{Phi-25dasympt}
it is found that the RWZ variable $\Phi_{\ss{S}}(u)$ and the KS variable
$Z_{\ss{S}}(u)$ yield the same quasinormal spectrum in five-dimensional
spacetimes as soon as one imposes the condition $
{\Phi^{\ss{in}}_{\ss{S}}}/{\sqrt{u}}-
{\mathcal{C}}_{\ss{S}}=0 $
at infinity ($ u\,\rightarrow 0$).

\subsubsection{Six and higher dimensions}

Following the same procedure as for $d=4$ and $d=5$ above,
Dirichlet boundary condition $Z^{\ss{in}}_{\ss{S}}(u=0)=0$ and
Eqs.~\eqref{Z2Phi2n} and \eqref{Phi-2ddasympt} yield
\begin{equation}
 (d-3)(d-5){\mathcal{C}}_{\ss{S}}(\mathfrak{w},\mathfrak{q})=0, 
\end{equation}
from what we conclude that for $(d-3)(d-5)\neq 0$ the quasinormal spectra
furnished by the master variables $Z_{\ss{S}}(u)$ and $\Phi_{\ss{S}}(u)$
are identical.

It is worth stressing here the relevance of the above results. They
allow us to use the most convenient gauge-invariant equations for each
specific case. For instance, whenever one has any kind of difficulty in
finding QN frequencies with a certain set of equations based, say, on the
KS gauge-invariant variables, one can try the other set of equations, based
on the RWZ variables. Moreover, some numerical methods require
Schr\"odinger-like equations such as in the case of the
time-evolution method used in the present work, as we will see in section
\ref{secNumericalQNM}.

\section{The quasinormal spectra: analytical results}
\label{qnmresults}

In this section we report on the procedure for calculating the QNM in some
asymptotic limits where results can be expressed in closed form.
In particular, the hydrodynamic limit of the QNM dispersion
relations are obtained analytically. A brief analysis of the results
is given for each sector with calculations done considering an arbitrary 
number of spacetime dimensions $d$. 
Other asymptotic regions of the QNM spectra such as large frequencies
and large spacetime dimensions are also analyzed.

\subsection{The hydrodynamic limit} \label{hydrolimit}

The hydrodynamic limit is the regime in which $\omega$ and $k$ are
sufficiently smaller than the Hawking temperature $T$, i.e., 
$\mathfrak{w},\,\mathfrak{q}\ll 1$. In such a regime it is possible to
express the solutions of the perturbation equations in the form of power
series in $\mathfrak{w}$ and $\mathfrak{q}$. By keeping just the lowest
order terms one finds the so-called hydrodynamic limit of the dispersion
relations ($\mathfrak{w}\rightarrow 0$, $\mathfrak{q}\rightarrow 0$). Such
a procedure is well known in the literature, and we do not reproduce it
here. The hydrodynamic limit of the dispersion relations to first order
approximation are known for some particular number of dimensions.
For instance, the vectorial and scalar sectors with $d=4,\,7$,
were treated in \cite{Herzog:2002fn,Herzog:2003ke}, and in five
dimensions the topic was explored in Refs.
\cite{Policastro:2002se,Policastro:2002tn,Kovtun:2005ev,Miranda:2008vb}.
This limit of the QNM spectra to second order approximation for $d=4,\,5,\,7$
has been studied in Refs.~\cite{Baier:2007ix,Natsuume:2007ty,Natsuume:2008iy}.
Here we show the results to first order approximation for all spacetime
dimensions and for all sectors of metric perturbations. We work with
the KS gauge-invariant variables and Dirichlet boundary condition at $u=0$.

\subsubsection{Tensorial perturbations}
\label{subsecPertTensor}

In the limit of small frequencies and small wavenumbers we find the
solution to Eq.~\eqref{fund-eq-tensor}, satisfying the
condition of representing ingoing waves at the horizon, as
\begin{equation}
Z^{\ss{in}}_{\ss{T}}(u)=\mathscr{C}_{\ss{T}}\,f^{i\mathfrak{w}/(d-1)}
[1+{\cal{O}}\left(\mathfrak{w}^{2}\right)],
\end{equation}
where $\mathscr{C}_{\ss{T}}$ is an arbitrary normalization constant.
Imposing the Dirichlet condition at anti-de Sitter boundary $u=0$, namely
$Z^{\ss{in}}_{\ss{T}}(0)=0$, and noting that $f(0)=1$, it follows that
there is no solution to  Eq. \eqref{fund-eq-tensor} satisfying the QNM
boundary conditions and being also a hydrodynamic QNM (i.e., satisfying
$|\mathfrak{w}|\ll1$ and $|\mathfrak{q}|\ll1$). The non-existence
of tensorial hydrodynamic QNM is compatible with the expectations from
hydrodynamics \cite{Kovtun:2005ev}.

\subsubsection{Vectorial perturbations}
\label{subsecPertVector}

The first order perturbative solution to Eq.~\eqref{fund-eq-vector}
satisfying the condition of representing an ingoing wave at the horizon is
given by
\begin{equation}
Z^{\ss{in}}_{\ss{V}}=\mathscr{C}_{\ss{V}}\,f^{i\mathfrak{w}/(d-1)}
\left[1-\frac{i\,\mathfrak{q}^{2}f}{(d-1)\,\mathfrak{w}}
+{\cal{O}}\left(\mathfrak{w}^{2}\right)\right],
\end{equation}
with $\mathscr{C}_{\ss{V}}$ being a normalization constant. The Dirichlet boundary
condition at infinity, $Z^{\ss{in}}_{\ss{V}}(0)=0$, implies the
following dispersion relation:
\begin{equation}\label{hidr}
\mathfrak{w} =
\frac{i}{d-1} \mathfrak{q}^{2}+{\cal{O}}\left(\mathfrak{q}^{3}\right).
\end{equation}
The dispersion relation \eqref{hidr} can be interpreted in terms of 
traveling waves in non-ideal fluids. In fact, it is expected from
hydrodynamics that a transversal momentum fluctuation presents a
shear mode, corresponding to a purely damped mode
with dispersion relation \cite{Fetter}
\begin{equation}\label{iDk}
\mathfrak{w} = i\frac{4\pi\,T}{d-1}\,D\, \mathfrak{q}^2,
\end{equation}
with $D$ being a diffusion constant carrying dimensions of length.
Therefore, the result in Eq.~\eqref{hidr} agrees with hydrodynamics
and the quasinormal frequency can be interpreted as the dispersion
relation for the shear mode, with diffusion constant
$D=1/4 \pi T$.

Finally, it is worth noticing that relation \eqref{hidr} holds for 
gravitational perturbations of plane-symmetric black holes in 
asymptotically AdS spacetimes of any dimension $d \geq 4$, and it is in 
agreement with previous results for $d=4$, $5$, $7$
(see \cite{Herzog:2002fn,Policastro:2002se,Kovtun:2005ev,Miranda:2008vb,
Mas:2007ng,kapusta:066017}).

\subsubsection{Scalar perturbations}
\label{subsecPertScalar}

Solving Eq.~\eqref{fund-eq-scalar} perturbatively in a power series in
$\mathfrak{w}$ and $\mathfrak{q}$ yields
\begin{equation}
Z^{\ss{in}}_{\ss{S}}=\mathscr{C}_{\ss{S}}\,f^{i\mathfrak{w}/(d-1)} 
\left\{\left[2 -2(d-2)\frac{\mathfrak{w}^{2}} {\mathfrak{q}^{2}}
-(d-3)\left(f-1\right)^2\right]+ \frac{4i\mathfrak{w}(d-3)f}{(d-1)}
+{\cal{O}} (\mathfrak{w}^{2} )\right\},
\end{equation}
with $\mathscr{C}_{\ss{S}}$ being an integration (normalization) constant. Imposing
the Dirichlet boundary condition at $u=0$ on $Z^{\ss{in}}_{\ss{S}}(u)$, and
taking into account we are working in the hydrodynamic limit, we obtain
\begin{equation}\label{hidr2}
\mathfrak{w} = \pm \frac{\mathfrak{q}}{\sqrt{d-2}}+\frac{(d-3)\, i}
{(d-2)(d-1)}\mathfrak{q}^{2}+ {\cal{O}}\left(\mathfrak{q}^{3}\right).
\end{equation}
In order to compare the above result \eqref{hidr2} to hydrodynamics we
first observe that for a conformal field theory the energy-momentum tensor
is traceless, so that the energy density $\varepsilon$ and the pressure $P$
of the dual plasma are related by $\varepsilon=(d-2)P$ and, consequently,
the speed of sound in the medium is $v_{s}=|\partial
P/\partial\varepsilon|^{1/2}=1/\sqrt{d-2\,}$. Thus, the expected dispersion
relation for the longitudinal momentum fluctuations, in the hydrodynamic
limit, must correspond to the sound wave mode \cite{Fetter}
 \begin{equation}\label{som}
\mathfrak{w}=\pm v_{s}\mathfrak{q}+\frac{4\pi\,i\,(d-3)T}{(d-2)(d-1)}\,D\,
\mathfrak{q}^{2}\, .
\end{equation}
The constant $D$ in Eq.~\eqref{som} is the same diffusion constant
appearing in Eq.~\eqref{iDk}. In fact, comparing Eqs.~\eqref{hidr2}
and \eqref{som} we find $D=1/4 \pi T$, agreeing with the value found
from the analysis of the hydrodynamic limit of vectorial perturbations.
This shows that the result given in Eq.~\eqref{hidr2} is consistent
with the expected result from hydrodynamics. Furthermore, this result 
is also in agreement with the previous results in the literature
for $d=4,\,5,\,7$ (see
\cite{Herzog:2003ke,Policastro:2002tn,Kovtun:2005ev,Natsuume:2007ty, 
Miranda:2008vb,Mas:2007ng,springer:086003}).

\subsection{Asymptotic analysis of the QNM}
\label{subsecAsym}

\subsubsection{Small wavenumbers, large frequencies}

There is an alternative analysis for large frequencies with finite
wavenumbers, namely $\mathfrak{w}\gg\mathfrak{q} $. To first order
approximation such a condition is equivalent to the asymptotic limit
$\mathfrak{q}\rightarrow 0$, as far as all the other parameters of the
model are kept fixed. That is to say, taking the limit $
\mathfrak{w} \rightarrow\infty$ with fixed $\mathfrak{q} $ yields the same
approximate equation as taking the limit $\mathfrak{q}\rightarrow 0$ with
finite $\mathfrak{w}$. For all of the perturbation equations
\eqref{fund-eq-tensor}, \eqref{fund-eq-vector} and \eqref{fund-eq-scalar}
with a little algebra we find 
\begin{equation}
Z''_{p}-\left[\frac{d-1-f}{uf}\right]Z'_{p}+
\frac{\mathfrak{w}^{2}}{f^{2}}Z_{p}=0,  \label{highw} 
\end{equation} 
where $p$ denotes the perturbative sector, as already
indicated. It is obvious that Eq. \eqref{highw} necessarily
imply in identical non-hydrodynamic quasinormal frequencies at
$\mathfrak{q}=0$ for all of the perturbation types. The same result was
also found in our numerical calculations, as it will be seen in the next
section (see Table \ref{tabq0}). With this result we conclude that the
dispersion relations for large frequencies are the same for all the three
perturbation sectors of a black brane, a result which was already
obtained by Nat\'ario and Schiappa \cite{Natario:2004jd} for the
Schwarzschild-AdS (Kottler) solution.

\subsubsection{Large number of spacetime dimensions}  
\label{largedim}

In this section we analyze the perturbation equations when the
number of spacetime dimensions is large,
namely $d\rightarrow\infty$ with finite $\mathfrak{w}$ and $\mathfrak{q}$.
For simplicity, in this analysis we consider the master equations for the
RWZ gauge-invariant variables (Eq.~\eqref{fund-eq-RWZ}), in which case the
analysis reduces to investigate the asymptotic form of the potentials
\eqref{pott}, \eqref{potv} and \eqref{pots} in the limit
$d \gg 4$. We thus find
\begin{equation}
 V_{\ss{T}}\rightarrow \frac{d^{2}}{4u^{2}}f\left(1
+u^{d-1}\right),
\end{equation}
\begin{equation}
 V_{\ss{V}}\rightarrow \frac{d^{2}}{4u^{2}}f\left(1
-3u^{d-1}\right),
\end{equation}
\begin{equation}
 V_{\ss{S}}\rightarrow \frac{d^{2}}{4u^{2}}f\left(1
+u^{d-1}\right).
\end{equation}
It is seen that in such a limit the tensorial and scalar potentials are the
same. Moreover, in the intervening region between the AdS
boundary and the horizon ($0<u<1$), the second term of the above
expressions within the parentheses tend
to zero in the limit $d\rightarrow\infty$, so that the potentials are
identical in this region. Moreover, the tensorial, scalar and
vectorial potentials approach the same values at the boundaries, namely 
$\lim_{u\rightarrow 0} V_p = d^{2}/4u^{2}$ and
$\lim_{u\rightarrow 1} V_p = 0$. These results suggest that the
QNM spectra of the three perturbation sectors for large $d$ are
identical. This is an important result because it shows the isospectrality
of the gravitational QNM of higher-dimensional AdS black holes.
Let us observe that this cannot be seen in our graphs because our
values of $d$ are not large enough when compared to the other parameters,
in particular $d\sim 4$ in our numerical results.

\section{Numerical results}
\label{secNumericalQNM}

\subsection{Methods}

We use two different methods to determine the gravitational QNM frequencies
of the black branes in the spacetime \eqref{fundo}. The first one is
a series expansion method \cite{Horowitz:1999jd,Berti:2009kk},
which reduces the problem to finding roots of a polynomial.
The second method employed in this work consists on a direct
time-evolution of the gravitational perturbations in these backgrounds
\cite{Wang:2000dt,Wang:2004bv}.

\subsubsection{Power series method}

The method developed by Horowitz and Hubeny \cite{Horowitz:1999jd}
consists in expanding the Fourier transformed perturbation
variables in power series of $u$ around the event horizon, $u=1$. The
condition of ingoing wave at the horizon is imposed on each
perturbation function. More specifically, the first step is to expand 
each of the functions $Z_p(u)$, $p=${\small{\it{T, V, S}}} in a Frobenius
series of the form $Z_p(u)=(1-u)^{i\mathfrak{w}/(d-1)}
\sum_j a_{j}(\mathfrak{w}, \mathfrak{q})(1-u)^{j}$.
The Dirichlet boundary condition at infinity is then imposed, and we obtain
an equation in the form of an infinite sum for
the coefficients, 
\begin{equation}\label{mod1}
\sum^{\infty}_{j=0}a_{j}(\mathfrak{w}, \mathfrak{q})=0,
\end{equation}
the roots of which yield the dispersion relation 
$\mathfrak{w}= \mathfrak{w}(\mathfrak{q})$.
During the calculation process, the infinite sum \eqref{mod1} is
truncated at a sufficiently large number of terms and then one finds
the roots of a polynomial in $\mathfrak{w}$. The accuracy
of the results is then verified through the relative variation between the
roots of two successive partial sums. The roots so obtained are the
quasinormal frequencies, which we write as
\begin{equation}\label{fre}
\mathfrak{w}= \mathfrak{w}_{\ss{R}} + i\; \mathfrak{w}_{\ss{I}}\, .
\end{equation}

Even though the method developed by Horowitz and Hubeny
\cite{Horowitz:1999jd} is well suited to large AdS black holes and black
branes, it is found that the convergence properties worsen for large
wavenumbers. Moreover, the capability of the Horowitz-Hubeny method in
finding the QN frequencies depends in an unclear way on the variables
chosen, on the considered region of the spectrum one seeks for solutions
and on the spacetime dimension. For instance, by using the master equation
for the KS variables, this method produced dispersion relations of
vectorial modes only for $\mathfrak{q}<4$. Then, by shifting to the
master equation for the RWZ variables we were
able to find satisfactory results for larger wavenumbers,
at least for $d=4,5$ and $6$. For higher spacetime
dimensions convergence problems occur for all perturbation sectors. In
particular, for the tensorial and scalar sectors via the master equations
with KS variables the numerical convergence problems of the series
solutions arise for dimensions larger than six ($d>6$), and higher
overtones ($n>1$ or $2$), even for intermediate wavenumber values
($\mathfrak{q} \sim 1$). Because of these convergence problems we used this
method to compute the dispersion relations for the first five quasinormal
modes for each perturbation sector, only for $d=4,\,5$ and $6$. In higher
dimensions we used a different method, a time-domain evolution method,
which allows one to read off the fundamental QNM for each sector, as seen
in the following, directly from the decay timescale and ringing frequency
of the signal.

\subsubsection{Time evolution method}
\label{t_evol-method}

The time evolution approach employed in the present work is based on a
characteristic initial value formulation of the perturbation wave
equations \cite{Price,Gundlach,Wang:2000dt}. The time-domain versions of
the equations \eqref{fund-eq-RWZ} are rewritten in terms of the
normalized light-cone variables $w =r_{h}\,t/R^{2}-\mathfrak{r}_{\ast}$ and
$v =r_{h}\,t/R^{2}+\mathfrak{r}_{\ast}$. The wave equations are integrated
numerically using the finite difference scheme introduced in
\cite{Wang:2004bv},
\begin{equation}
\begin{split}
\left[1-\frac{\Delta^{2}}{16}V_{p}(S)\right]\Phi_{p}(N)=&\,
\Phi_{p}(E) + \Phi_{p}(W)-\Phi_{p}(S)\\
&-\frac{\Delta^{2}}{16} \left[V_{p}(S)\Phi_{p}(S)
+ V_{p}(E)\Phi_{p}(E) + V_{p}(W)\Phi_{p}(W)\right] \,\,.   
\end{split}
\end{equation}   

The scalar, vectorial and tensorial sectors are indexed by $p$. The
points $N$, $S$, $W$ and $E$ are defined as: $N = (w + \Delta,v+ \Delta)$,
$W = (w + \Delta, v)$, $E = (w, v + \Delta)$ and $S =(w,v)$. The
discretization step $\Delta$ is a function of the grid size and number of
point in the grid \cite{Wang:2000dt}.

Initial data are specified on the null surfaces $w = w_{0}$ and $v =
v_{0}$. Since the behavior of the integrated wave functions is largely
insensitive to the choice of initial data (which was empirically verified
in the present scenario), we set $\Phi_{p}(w,v=v_{0})=0$ and use a Gaussian
pulse as initial perturbation  $\Phi_{p}(w=w_0,v)=\exp[-(v -
v_{c})^2/2\sigma^2 ]$. After the integration is completed, the values of
$\Phi_{p}$ on selected curves are extracted. The quasinormal fundamental
frequency can (usually) be accurately estimated from the data.

The algorithm precision depends on the number of points and size of
the discretized grid. One basic requirement for the method is the
convergence of the code with respect to the variation of the number
of grid points. It was observed in our numerical experiments that the
convergence rate varied with the parameters of the system, with the
worse performance in the small $\mathfrak{q}$ limit for the tensorial and
scalar sectors. Since in this limit the power series method is
usually reliable, and the concordance of both methods is very good in
a wide range of parameter space, it is accurate to say that the methods
employed in this work are complementary.

In the following we show some numerical results and analyze in some
detail the dispersion relations of the fundamental mode, i.e., the QNM
with the smallest imaginary part of the frequency, for each
perturbation sector. As we see below, for the vectorial and
scalar sectors, the fundamental modes in the low wavenumber regime are in
fact hydrodynamic QNM, so denominated because they present a
characteristic behavior in the hydrodynamic limit, $\mathfrak{w}
\rightarrow 0$ when $\mathfrak{q} \rightarrow 0$. The basic motivation
for studying the hydrodynamic modes is because they are important
modes in connection to the AdS/CFT correspondence, since they furnish
(for low $\mathfrak{q}$ values) the thermalization time in the conformal
field theory at AdS spatial infinity.

\subsection{Numerical results for the tensorial QNM}
\label{subsecMQNGT}

In Table \ref{tab1} we list the values obtained for the QN
frequencies of the first five modes (overtone numbers $n=0,1,...,4$),
with $\mathfrak{q}=0$ in five and six spacetime dimensions. From the
results shown in this table, it is seen that the
fundamental equation for the tensorial gravitational perturbations
\eqref{fund-eq-tensor} for $d=5$ and the use of the Horowitz-Hubeny
method reproduce the results in the literature \cite{Starinets:2002br}
with very good accuracy. For six dimensions, we have not found any
previous result in the literature for comparison.

\TABLE{
\begin{tabular}{ccccc}
\hline 
\hline & \multicolumn{2}{c}{$d=5$} & 
\multicolumn{2}{c}{$d=6$}
\\ \hline
$\;\;n\;\;$ & $\quad\;\;\mathfrak{w}_{\ss{R}}\quad\;\;$ &
$\quad\;\;\mathfrak{w}_{\ss{I}}\quad\;\;$ & $\quad\;\;
\mathfrak{w}_{\ss{R}}\quad\;\;$ &
$\quad\;\;\mathfrak{w}_{\ss{I}}\quad\;\;$ \\
\hline 
0 & 3.11945 & 2.74668 & 4.13591  & 2.69339 \\
1 & 5.16952 & 4.76357 & 6.60919  & 4.45349  \\
2 & 7.18793 & 6.76957 & 9.02574  & 6.19321 \\
3 & 9.19720 & 8.77248 & 11.4241  & 7.92685 \\
4 & 11.2027 & 10.7742 & 13.8136 & 9.65854 \\ \hline\hline
\end{tabular}
\caption{\small{The first five tensorial QNM with zero wavenumber for
a plane-symmetric AdS black hole (black brane) in five and six
spacetime dimensions.}}\label{tab1}}

The dispersion relations for the fundamental tensorial QNM for
$d=5,\,6,\,...,\,10$ are shown in Fig.~\ref{figten1vard}. The
Horowitz-Hubeny method yielded the results for $d=5,\,6$
and for small wavenumbers in $d=7$. The time-evolution method was used to
compute the dispersion relations for the other dimensions and for
large wavenumbers in $d=7$. We observe some convergence problems
for small values of $\mathfrak{q}$ and $d>7$, as it is
apparent from the dispersion relation curves for the imaginary part of the
frequency (see the right panel of Fig.~\ref{figten1vard}). 

\FIGURE{
\centering\epsfig{file=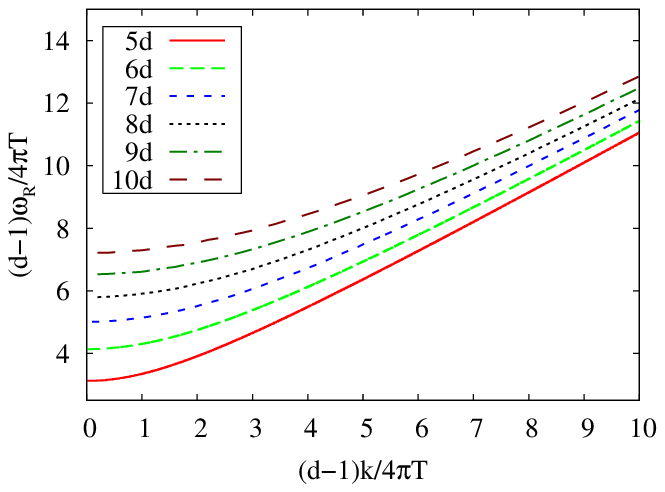,
width=5.15cm, angle=270}
\centering\epsfig{file=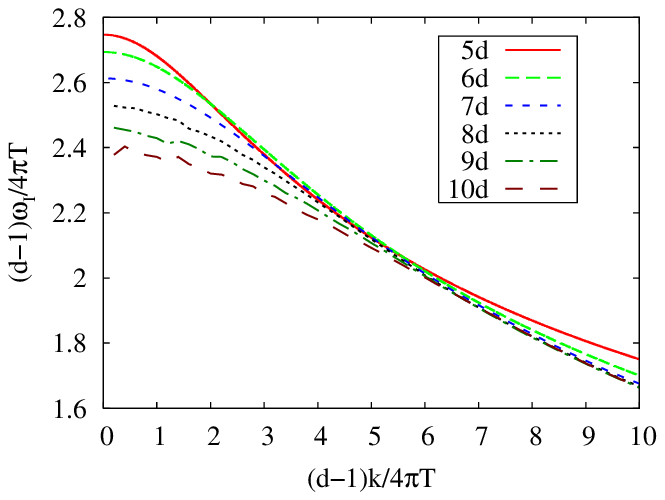,
width=5.15cm, angle=270}
\caption{Dispersion relations for the first tensorial QNM of AdS
black branes in several dimensions, $d=5,\,6,\,...,\,10$, as indicated.
The small wiggles in the curves of the imaginary parts of the
frequencies for larger values of $d$ indicate numerical
convergence problems (see the text).} \label{figten1vard}}

It is also apparent in Fig.~\ref{figten1vard} that the real and imaginary
parts of the QNM frequencies present an overall behavior that appears to be
independent of the number of spacetime dimensions. Dispersion relations
$\mathfrak{w}_{\ss{R}}(\mathfrak{q})$ approaching straight lines and
dispersion relations $\mathfrak{w}_{\ss{I}}(\mathfrak{q})$ approaching zero
as the wavenumber $\mathfrak{q}$ increases. The same feature is also shown
for higher overtones in Fig. \ref{figten6d}. However, the larger the
overtone index $n$ and/or the larger the number of dimensions, the faster
the frequency $\mathfrak{w}_{\ss{I}}$ approaches zero. This is in agreement
with the results by Festuccia and Liu \cite{Festuccia:2008zx} (see also
Ref.~\cite{Morgan:2009vg}). In fact, it was shown in
\cite{Festuccia:2008zx} by means of analytical methods, and confirmed in
\cite{Morgan:2009vg} through numerical methods, that the dispersion
relations for large wavenumbers are given approximately by
$\mathfrak{w}_{\ss{R}}=\mathfrak{q}+\alpha_{\ss{R}}\mathfrak{q}^{-\beta}$
and $\mathfrak{w}_{\ss{I}}=\alpha_{\ss{I}}\mathfrak{q}^{-\beta}$, where
$\alpha_{\ss{R},\ss{I}}$ are parameters depending on the number of
dimensions and overtone index, and $\beta= (d-3)/(d+1)$.

\FIGURE{
\centering\epsfig{file=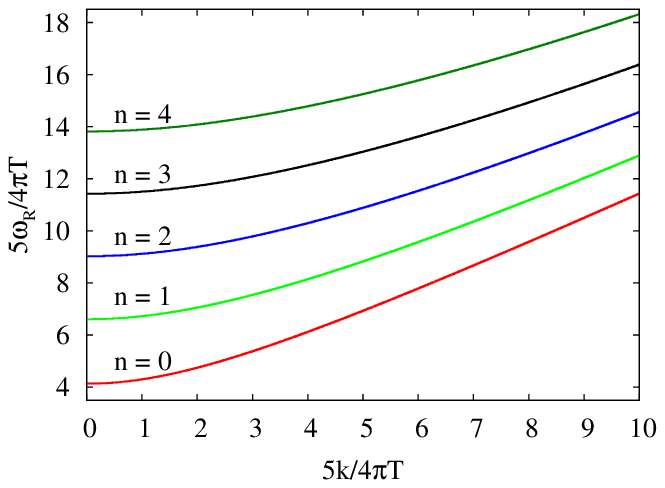, width=5.15cm,angle=270}
\centering\epsfig{file=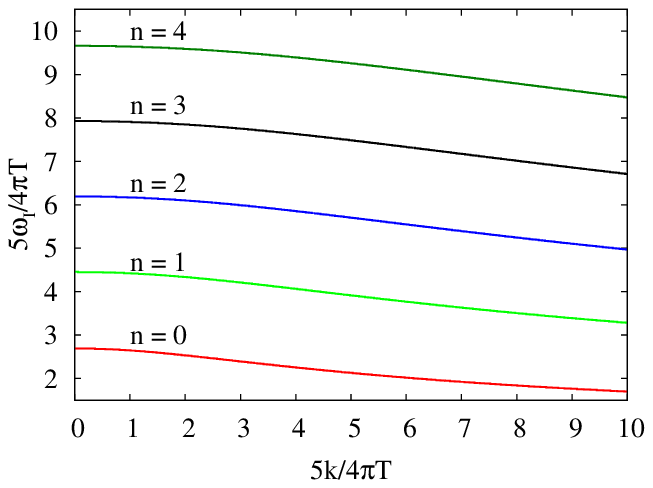, width=5.15cm,angle=270}
\caption{Dispersion relations for the first five tensorial QNM
of AdS black branes in $d=6$.}
\label{figten6d}}

In Fig. \ref{figten6d} we have the dispersion relations
for the first five tensorial modes in six dimensions where we can see
that for a finite fixed temperature, the imaginary parts of
the frequencies decrease with the wavenumber value, while the
real parts increase with the wavenumber.
The behavior is qualitatively similar for $d=5$.

\FIGURE{
\centering\epsfig{file=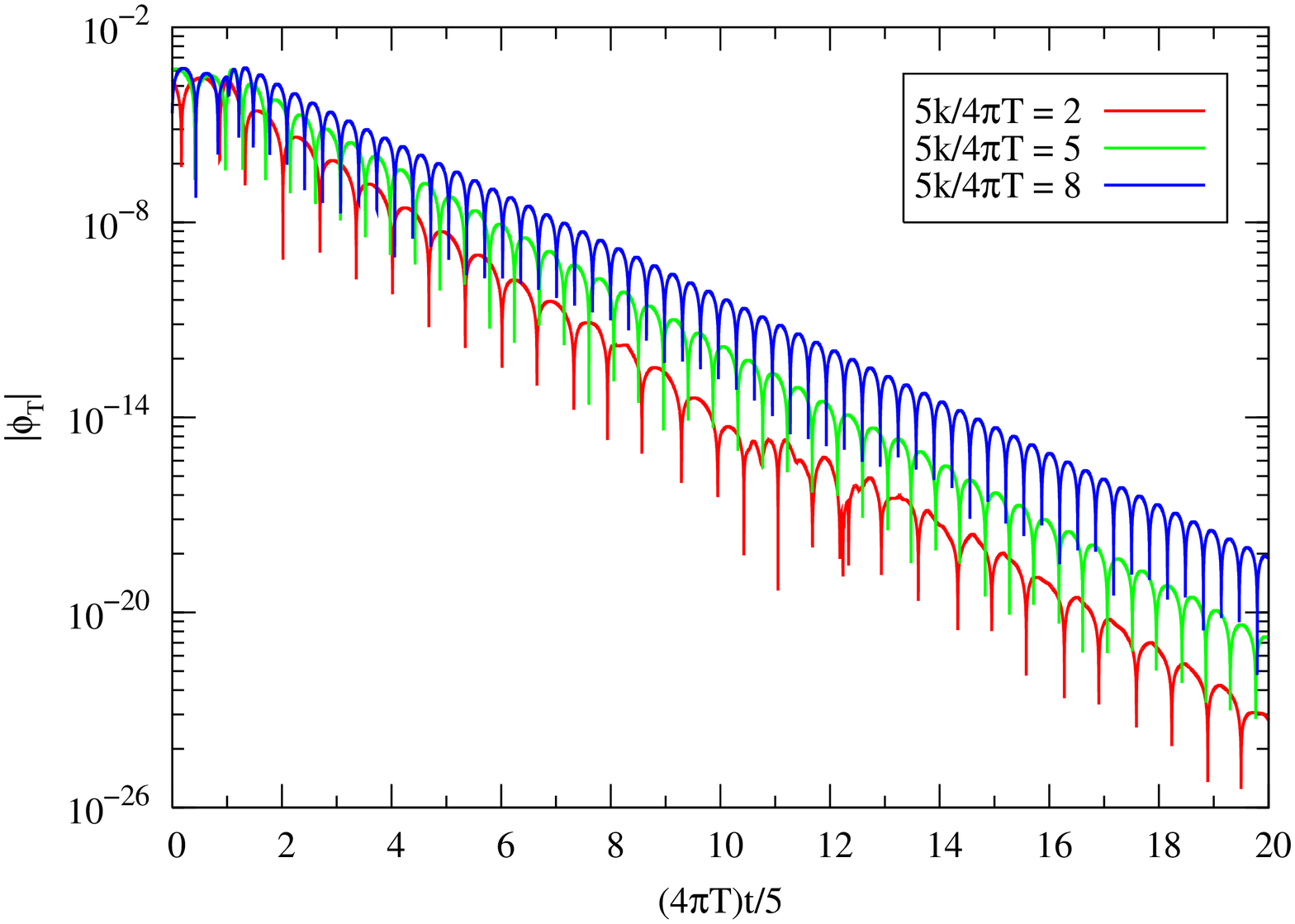,
width=7.15cm, angle=270}
\caption{Time-evolution profiles of the tensorial mode
perturbations for selected values of $\mathfrak{q}$ in $d=6$ (from top to
bottom, for $\mathfrak{q}=8,\,5,\,2)$.}
\label{timevolution1}}

Time-evolution profiles of the tensorial perturbations can be
seen in Fig. \ref{timevolution1}. The oscillatory decay, which is
characteristic of a QNM with $\mathfrak{w}_{\ss{R}}\neq 0$, dominates the
intermediate- and late-time behavior of the wave function $\Phi_{\ss{T}}$.
In contrast with what happens in asymptotically flat spacetimes, our
numerical results for all perturbation sectors show no sign of a power-law
tail at late stages, in agreement with
earlier predictions\cite{Horowitz:1999jd,Ching:1995tj}.

As expected from the study of the hydrodynamic limit of tensorial quasinormal
modes given in section \ref{subsecPertTensor}, the numerical analysis of
Eq.~\eqref{fund-eq-tensor} based on the Horowitz-Hubeny approach
found no QNM satisfying simultaneously the conditions
$\mathfrak{w}\ll 1$ and $\mathfrak{q}\ll 1$.

\subsection{Numerical results for the vectorial QNM}
\label{subsecMQNGV}

The frequencies for the hydrodynamic and for the first five
non-hydrodynamic vectorial QNM are shown in Table \ref{tab2a}, where we
have set $d=5,6$ and $\mathfrak{q}=2$. The results for five dimensions
coincide with those of Ref.~\cite{Kovtun:2005ev}. As far as we are aware
of, there are no data in the literature for comparison to the results shown
in Table \ref{tab2a} for six dimensions.

\TABLE{
\begin{tabular}{ccccc}
\hline 
\hline & \multicolumn{2}{c}{$d=5$} & 
\multicolumn{2}{c}{$d=6$}
\\ \hline
$\;\;n\;\;$ & $\quad\;\;\mathfrak{w}_{\ss{R}}\quad\;\;$ &
$\quad\;\;\mathfrak{w}_{\ss{I}}\quad\;\;$ & $\quad\;\;
\mathfrak{w}_{\ss{R}}\quad\;\;$ &
$\quad\;\;\mathfrak{w}_{\ss{I}}\quad\;\;$ \\
\hline 
0 & 0 & 1.19612 &   0  & 0.87233\\
1 & 3.51823 & 2.58319 & 4.51460 & 2.57492 \\
2 & 5.46616 & 4.66081 & 6.88034 & 4.37633 \\
3 & 7.43187 & 6.69069 & 9.24255 & 6.13420 \\
4 & 9.40729 & 8.70698 & 11.6070 & 7.87837 \\
5 & 11.3889 & 10.7175 & 13.9739 & 9.61647 \\
\hline\hline
\end{tabular}
\caption{Some data for the frequencies of the vectorial
gravitational QNM of five and six dimensions for $\mathfrak{q}=2$.}
\label{tab2a}}

In Fig.~\ref{figvet1vard} we plot the hydrodynamic
and the first non-hydrodynamic vectorial QNM for several dimensions,
$d=4,\,5,\,...,\,10$, as indicated (recall that the hydrodynamic modes are
those for which the frequencies $\mathfrak{w} (\mathfrak{q})$ vanish when
$\mathfrak{q}\rightarrow 0$). The numerical results confirm the fact that
the hydrodynamic vectorial QNM are purely damped modes, and so the
dispersion relation curves have just the imaginary parts, as shown in the
right panel of Fig.~\ref{figvet1vard}. Since we could not find the QNM
using the Horowitz-Hubeny method for $d>7$, the dispersion relations for
$d=8,\,9,\,10$ were obtained through a time-domain evolution method, as
described in section \ref{t_evol-method}, so the curves for the
non-hydrodynamic QNM at low values of $\mathfrak{q}$ could not be found,
because the hydrodynamic mode dominates in that region. This is clearly
seen in the left panel of Fig.~\ref{figvet1vard}, where the dispersion
relation curves for the first non-hydrodynamic mode at small values of
$\mathfrak{q}$ and $d=8,\,9,\,10$ are missing. As a matter of fact, the
curves for $d=7$ were obtained by joining the results from both of the
numerical methods used here. We see that the dispersion relations for
$d=5,\,6,\,7$ have all the same behavior. A slight difference is observed
in the curve $\mathfrak{w}_R\times\mathfrak{q}$ for $d=4$ (see the lowest
curve of the left panel in Fig.~\ref{figvet1vard}) in which it is seen a
``knee'' around $\mathfrak{q}\simeq 2$. This local minimum in the real part
of the frequency is present in all of the gravitational vectorial modes of
the four-dimensional black brane \cite{Miranda:2008vb}. As seen from
Fig.~\ref{figvet1vard}, such a local minimum disappears in higher
dimensions.

\FIGURE{
\centering\epsfig{file=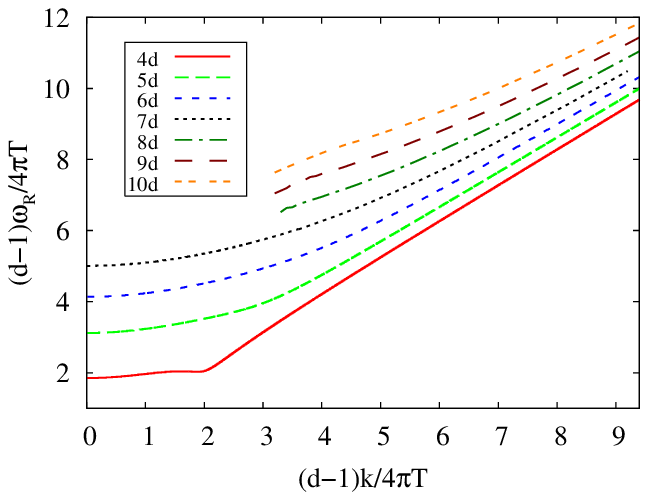,width=5.15cm, angle=270}
\centering\epsfig{file=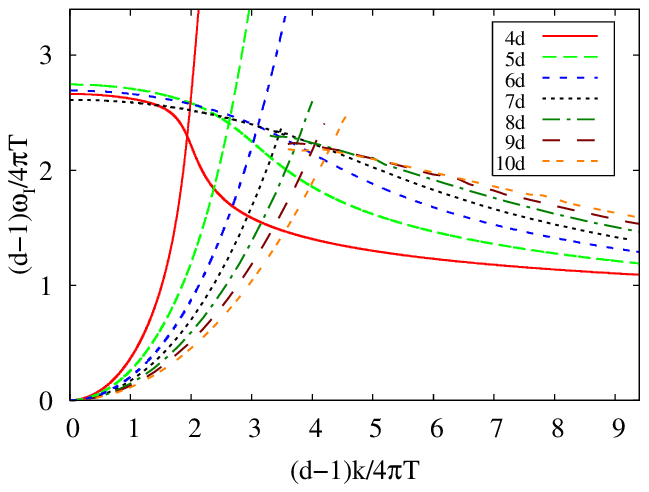,width=5.15cm, angle=270}
\caption{Dispersion relations for the first non-hydrodynamic
vectorial QNM of the AdS black brane in several dimensions,
$d=4,\,5,...,\,10$. The curves for the hydrodynamic modes
are also shown in the right panel for
$0\leq\mathfrak{q}\lesssim 4$.}
\label{figvet1vard}}

\FIGURE{
\centering\epsfig{file=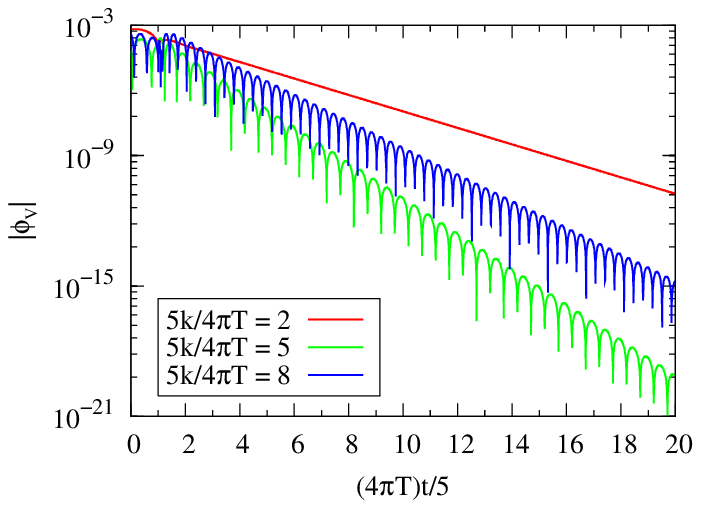, width=5.15cm, angle=270}
\centering\epsfig{file=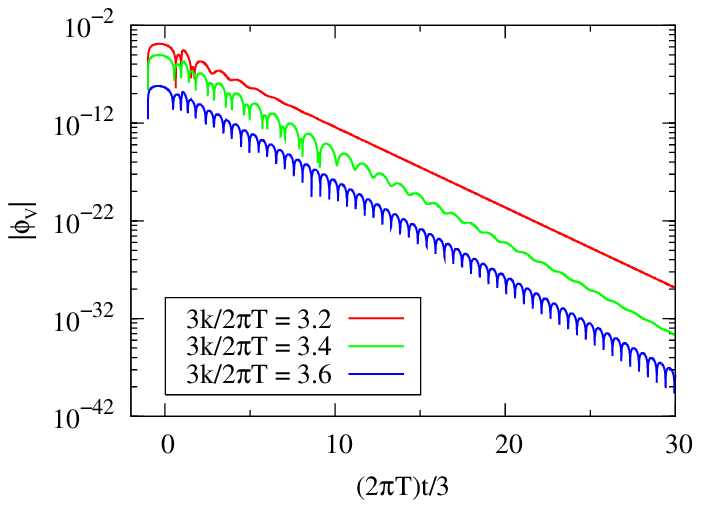, width=5.15cm, angle=270}
\caption{Time-evolution profiles of the vectorial mode
perturbations for $d=6$ (left panel) and $d=7$ (right panel). In the
left panel we plot (from top to bottom) the curves for
$\mathfrak{q}=2,\,8,\,5$ wavenumbers. The transition from
non-oscillatory ($\mathfrak{q}=2$) to oscillatory late-time decay
($\mathfrak{q}=5,\,8$) is clearly seen.  In the right panel, from top
to bottom, we see the curves for $\mathfrak{q}=3.2,\,3.4,\,3.6\,$. The
mode mixing is apparent.}
\label{timevolution2}}

Typical time-domain evolution profiles of the vectorial QNM are presented
in Fig.~\ref{timevolution2}. The transition from the hydrodynamic
shear-mode regime of perturbations to the ordinary-QNM regime appears in
the time evolution of $\Phi_{\ss{V}}$ as a transition from non-oscillatory
(for small $\mathfrak{q}$) to oscillatory (large $\mathfrak{q}$) late-time
decay. The left panel of Fig. \ref{timevolution2} exploits exactly this
feature of the vectorial gravitational QNM in $d=6$: non-oscillatory
time-evolution for $\mathfrak{q}=2$ and oscillatory time-evolution for
$\mathfrak{q}=5,\,8$. This transition is important to the CFT side of the
AdS/CFT correspondence, since it is interpreted as the
hydrodynamic-to-collisionless crossover which is expected to arise in
generic systems \cite{forster,Herzog:2007ij}. The right panel of
Fig.~\ref{timevolution2}, now for seven-dimensional black branes, shows the
mode mixing as seen in the two upper curves of such a figure. The choice of
values for $d$ and $\mathfrak{q}$ was made with the aim of showing not only
the transition from a non-oscillatory to an oscillatory regime, but also to
exploit the mode mixing feature of the QNM. This feature can be understood
by assuming that for intermediate times an oscillatory mode dominates,
while for later times a non-oscillatory mode dominates. In such a
situation, the identification of a specific QNM frequency from the
numerical data is very difficult. This explains the missing parts of the
dispersion relation curves for $d=7,\,8,\,9,\,10$ in
Fig.~\ref{figvet1vard}.

The graphs in Fig.~\ref{figvet6d} show the first five vectorial
non-hydrodynamic quasinormal modes for the six-di\-mensional black branes.
The hydrodynamic QNM and the shear mode of equation \eqref{hidr}
are also shown in the right panel of this figure. We can see that the
behavior of the non-hydrodynamic modes is very similar to
the tensorial QNM (cf. Fig. \ref{figten6d}).
Similarly to the tensorial sector, the higher overtone
modes ($n>1$) follow the overall behavior of the first non-hydrodynamic
QNM. Among the most prominent differences between vectorial and tensorial
QNM, we see that the imaginary parts of the frequencies approach zero
with growing $\mathfrak{q}$ faster in the vectorial case than in the
tensorial case, which means that the thermalization time in the dual
CFT is dominated by the vectorial perturbations in comparison to the
corresponding tensorial modes.

\FIGURE{
\centering\epsfig{file=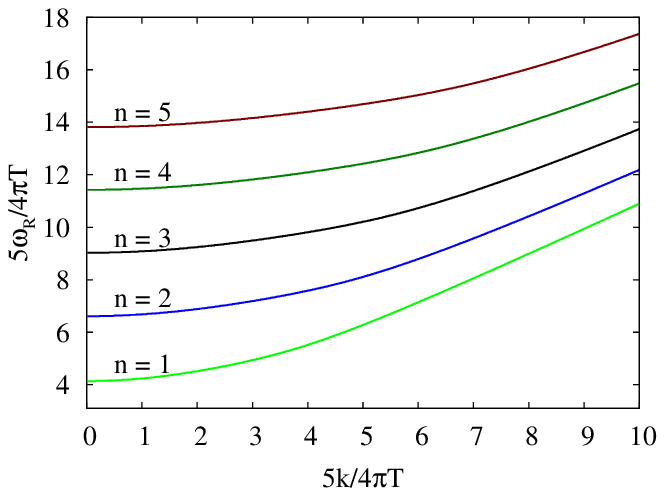,width=5.15cm, angle=270}
\centering\epsfig{file=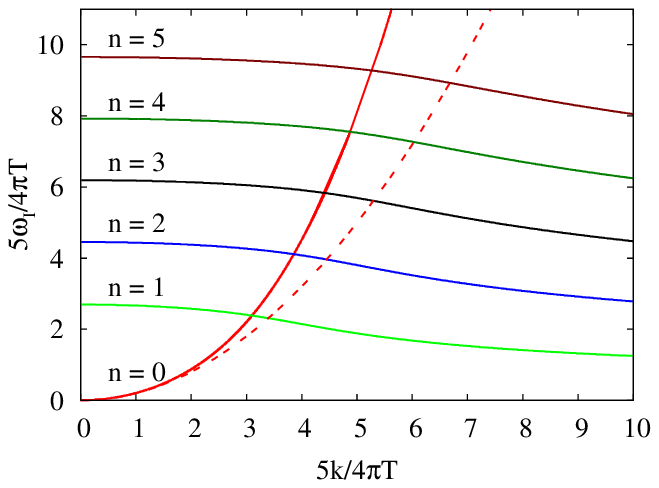,width=5.15cm, angle=270}
\caption{Dispersion relations for the hydrodynamic mode and for
the first five vectorial non-hydrodynamic QNM of the AdS black brane in six
dimensions. The dashed line in the right panel corresponds to the shear
mode of equation \eqref{hidr}.}
\label{figvet6d}}

The numerical results shown in Figs.~\ref{figvet1vard} and \ref{figvet6d}
agree with the analytical relations for small frequencies and wavenumbers
(cf. right panel of Fig.~\ref{figvet6d}), what corresponds to the
hydrodynamic limit of the QNM spectrum. The behavior of
the hydrodynamic mode is very important for the interpretation of
quasinormal modes in terms of the AdS/CFT correspondence since it furnishes
the value of the diffusion constant $D$ and dominates the thermalization
timescale of the perturbations for small wavenumbers. The study of the
thermalization time is one of the subjects of section \ref{subsecAdSCFT}.

\subsection{Numerical results for the scalar QNM}
\label{subsecMQNGE}

In Table \ref{tab3a} we list the hydrodynamic and the first five
non-hydrodynamic scalar QNM of the plane-symmetric AdS black holes in five
and six spacetime dimensions, fixing $\mathfrak{q}=2$. Again, our results
agree with those of Ref.~\cite{Kovtun:2005ev} for the five-dimensional
case, while in the six-dimensional case we have not found similar data in
the literature for comparison.

\TABLE{
\begin{tabular}{ccccc}
\hline 
\hline & \multicolumn{2}{c}{$d=5$} & 
\multicolumn{2}{c}{$d=6$}
\\ \hline
$\;\;n\;\;$ & $\quad\;\;\mathfrak{w}_{\ss{R}}\quad\;\;$ &
$\quad\;\;\mathfrak{w}_{\ss{I}}\quad\;\;$ & $\quad\;\;
\mathfrak{w}_{\ss{R}}\quad\;\;$ &
$\quad\;\;\mathfrak{w}_{\ss{I}}\quad\;\;$ \\
\hline 
0 & 1.48286 & 0.57256 & 1.19804 & 0.58389 \\
1 & 3.46702 & 2.68602 & 4.40613 & 2.62487 \\
2 & 5.41108 & 4.71412 & 6.79949 & 4.40716 \\
3 & 7.37878 & 6.72773 & 9.17633 & 6.15701 \\
4 & 9.35747 & 8.73596 & 11.5503 & 7.89671 \\
5 & 11.3422 & 10.7416 & 13.9214 & 9.63703 \\
\hline\hline
\end{tabular}
\caption{The frequencies of the hydrodynamic and the first five
non-hydrodynamic scalar QNM in five- and six-dimensional
spacetimes, with $\mathfrak{q}=2$. }
\label{tab3a}}

We have used the Horowitz-Hubeny method to obtain
the complete dispersion relations of the dominant scalar
quasinormal modes for $d=4,\,5,$ and $6$. These results are shown
in the graphics of Fig.~\ref{figscalarhydro}. The fundamental
QNM of the scalar perturbations is in fact the hydrodynamic
(sound-wave) mode. For large wavenumber values, the behavior of
the real and imaginary parts of the frequency is similar to that of the
tensorial and vectorial sectors, but it is quite different for small values
of $\mathfrak{q}$. Such a difference can be attributed to fact that the
dominant scalar QNM in the regime of small wavenumbers is a
hydrodynamic mode with nonvanishing real part.

\FIGURE{
\centering\epsfig{file=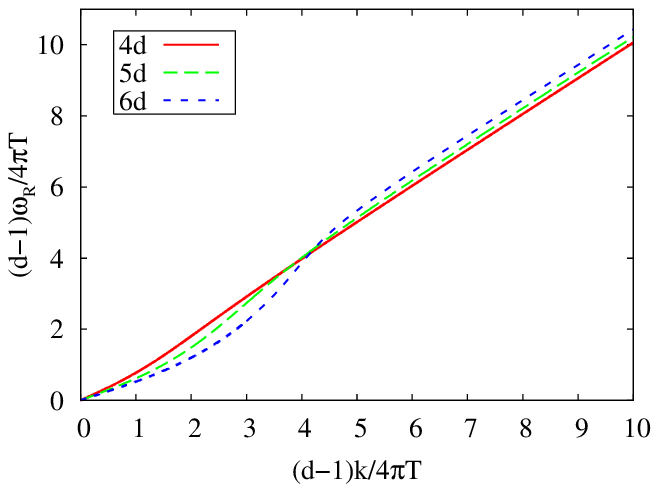,width=5.25cm, angle=270}
\centering\epsfig{file=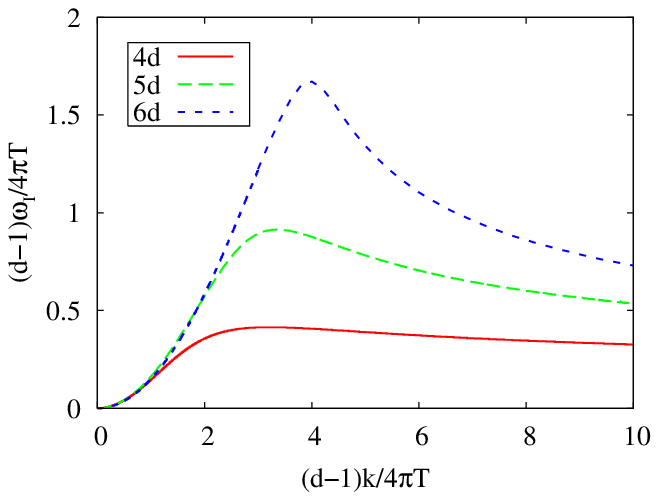,width=5.25cm, angle=270}
\caption{Dispersion relations for the dominant scalar QNM (the
hydrodynamic mode) of the plane-symmetric AdS black hole
in $d=4,\,5,\,6$ dimensions, as indicated.}
\label{figscalarhydro}}

Fig.~\ref{figscalarhydro2} displays the numerical results for the
QN frequencies of scalar gravitational perturbations for $d=7,\,...,\,10$
spacetime dimensions. In such cases, the dispersion relations
were obtained by the time-domain evolution
method. This method in general gives us information only on
the dominant (lowest-$\mathfrak{w}_{\ss{I}}$) mode, but as apparent in
Fig.~\ref{figscalarhydro2} we were able to obtain the dispersion relations
of the fundamental and the first excited modes in the region
$4\lesssim\mathfrak{q}\lesssim 5$.
The results shown in such figure indicate the
existence of a critical wavenumber value from which on the first
non-hydrodynamic QNM is the dominant mode of the scalar perturbations.
Such a result appears in the form of a gap in the real part of the
frequencies (left panel in Fig. \ref{figscalarhydro2}) and as a
crossing of two curves $\mathfrak{w}_{\ss{I}}(\mathfrak{q})$ in the
right panel of Fig. \ref{figscalarhydro2}. This behavior of the scalar QNM
in higher-dimensional spacetimes ($d>6$) is completely
different from what happens in $d=4$, $5$ and $6$ dimensions, where the
hydrodynamic scalar QNM dominates all of the spectrum.
A similar behavior with two concurrent dominant
modes in some region of the spectrum was also found for the
scalar-type gravitational perturbations of an asymptotically
flat black string in $d=5,6$ and $7$ dimensions \cite{konoplya:084012}.
However, in contrast with the black string
case of Ref. \cite{konoplya:084012}, we do not find
purely damped modes in this sector of the gravitational perturbations.

\FIGURE{
\centering\epsfig{file=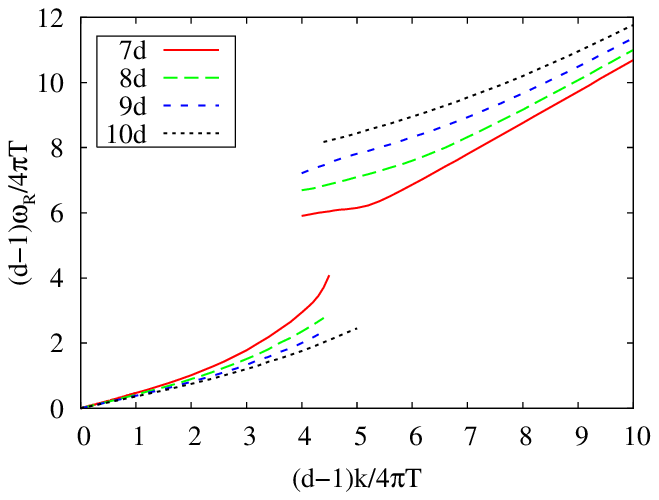,width=5.25cm, angle=270}
\centering\epsfig{file=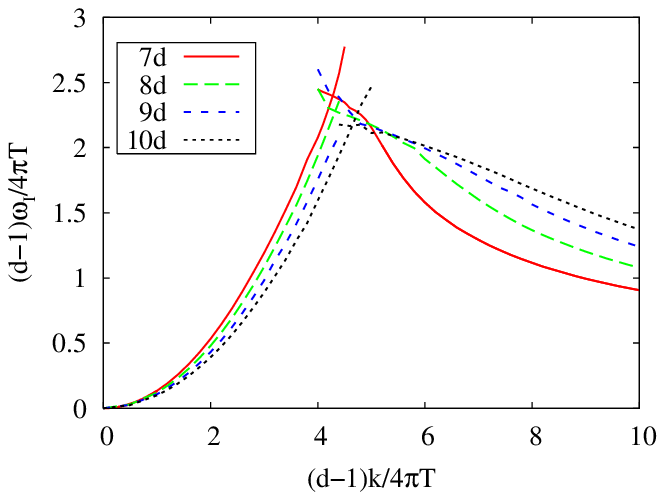,width=5.25cm, angle=270}
\caption{Dispersion relations for the dominant scalar QNM in
$d=7,\,8,\,9$ and $10$ dimensions, as indicated. Exceptionally in the
region $4\lesssim\mathfrak{q}\lesssim 5$, we present both the
fundamental and the first excited modes obtained from the
time-evolution method. Notice that the hydrodynamic QNM dominates in
the regime of small $\mathfrak{q}$, while the first non-hydrodynamic
mode is the dominant mode in the high-$\mathfrak{q}$
regime.}\label{figscalarhydro2}}

A few samples of the time-domain evolution profile of the scalar
perturbations are shown in Fig. \ref{timevolution3}. The oscillatory
decay, which is characteristic of a QNM with
$\mathfrak{w}_{\ss{R}}\neq 0$, dominates the intermediate- and
late-time behavior of the wave function $\Phi_{\ss{S}}$. Again we do
not see any power-law tail at late stages.

\FIGURE{
\centering\epsfig{file=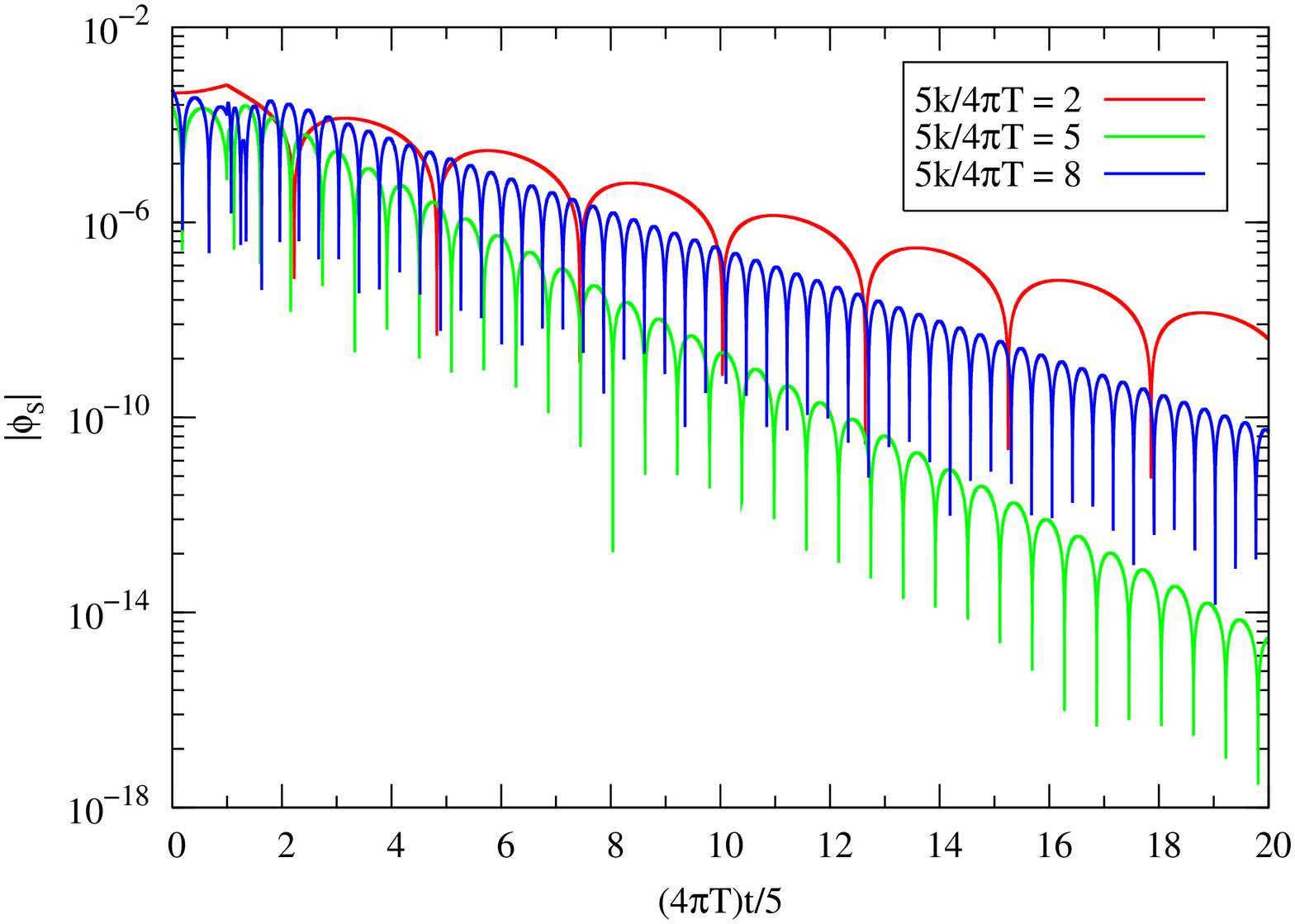,width=7.15cm, angle=270}
\caption{Time-evolution profiles of the scalar mode perturbation for
selected values of $\mathfrak{q}$ in $d=6$ (from top to bottom, the
curves are for $\mathfrak{q}=2,\,8,\,5$, respectively).}
\label{timevolution3}}

\FIGURE{
\centering\epsfig{file=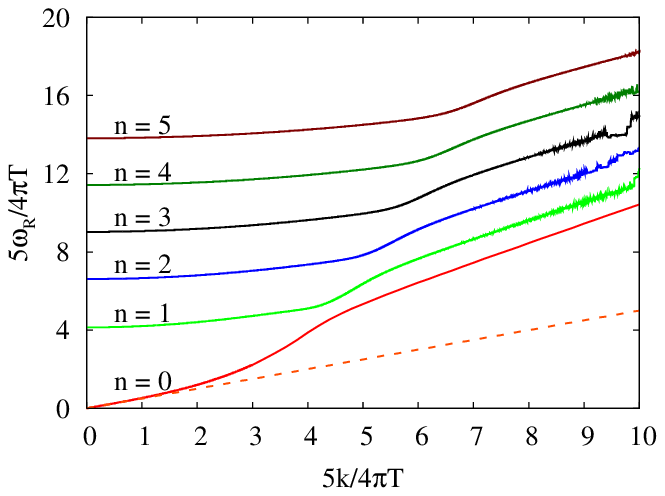,angle=270}
\centering\epsfig{file=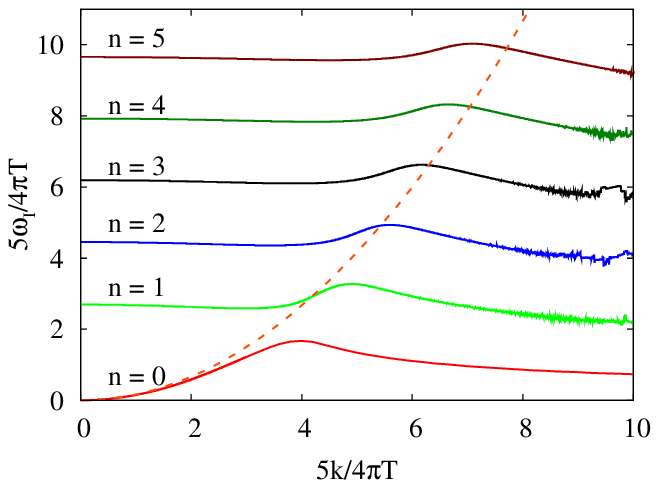,angle=270}
\caption{Dispersion relations for the hydrodynamic mode ($n=0$) and
for the first five scalar non-hydrodynamic QNM in six dimensions. The
dashed lines correspond to the sound-wave mode of equation
\eqref{hidr2}. The wiggles in the plots are due to numerical errors in
the Horowitz-Hubeny method.}
\label{figscal6d}}

The dispersion relation for the first five non-hydrodynamic scalar
QNM in six-di\-mensional spacetime have a behavior very similar to
the hydrodynamic mode and are shown in
Fig.~\ref{figscal6d}. The sound-wave mode is also plotted
in these graphs (the dashed line in each panel). 
We again can see that the
numerical results agree with the analytical results of equation
\eqref{hidr2} (dashed lines) in the limit
$\mathfrak{w},\mathfrak{q}\ll 1$. It is also seen that the 
peak (maximum) in the imaginary part of the frequency is smoothed out
as the overtone number $n$ increases, and the position of the
peak is displaced towards higher $\mathfrak{q}$-values. Such a
behavior is similar to the $d=5$ case. However, it is important to
note that the peaks for higher overtones ($n\ge 1$) in $d=4$
are located at the origin, $\mathfrak{q}=0$ (cf.
Ref.~\cite{Miranda:2008vb}). Aside this fact, it seems that the overall
profile of the dispersion relations does not strongly depend on
the number of spacetime dimensions, at least for
$d=4,\, 5$ and $6$. For scalar perturbations, the Horowitz-Hubeny method
presented numerical problems also in six dimensions, as seen in the
wiggling curves for high overtones and
high values of the wavenumber $\mathfrak{q}$ in Fig.~\ref{figscal6d}.
As far as we could check, the numerical error settles in as
$\mathfrak{q}$ grows and the convergence of the method becomes very
sensitive to the numerical precision. Even with these convergence
problems, the data in Fig.~\ref{figscal6d} is good enough to conclude
that, for higher values of $\mathfrak{q}$, the dispersion
relations for the scalar QNM of the AdS black branes behave in a
similar way as the vectorial and tensorial QNM studied above (cf.
Figs.~\ref{figten6d} and \ref{figvet6d}).

\subsection{More comments on the numerical results}

The QNM spectra give us also information about the stability of the black
hole background spacetime. Following this approach, Kodama and Ishibashi
\cite{Kodama:2003ck} have proved that black branes are stable against
tensorial and vectorial perturbations in all dimensions, and against
scalar perturbations in $d=4$. The stability of higher-dimensional black
branes against scalar perturbations from the analytical point of view is
still an open problem.
Our results for the tensorial and vectorial sectors are consistent
with the results by Kodama and Ishibashi \cite{Kodama:2003ck} and for
the scalar sector we have not found unstable modes in $d=4$, as
expected. Moreover, we have not found frequencies with negative
imaginary parts for small and intermediary wavenumbers $\mathfrak{q}$
in $d\geq5$ dimensions. In order to investigate a possible instability
for large wavenumbers we can consider the asymptotic behavior of the
dispersion relations. In \cite{Morgan:2009vg} it was observed that the
analytical prediction by Festuccia and Liu \cite{Festuccia:2008zx}
describes quantitatively the numerical results if multiplied by a real
function which depends only on the overtone $n$. Such results
show that the imaginary parts of the QN frequencies are positive for
all dimensions, which strongly suggests that these asymptotically AdS black
branes are stable against general gravitational perturbations.

In connection with the stability problem, a word of caution
should be given here about the geometric interpretation of the background
spacetime considered in this work. The plane-symmetric asymptotically
AdS black holes (black branes) should not be confused with other
higher-dimensional extended black objects appearing in the
literature. In particular, as we have shown above, the black-brane
spacetime \eqref{fundo} does not present any kind of
gravitational instability, independent of the parameter values.
This is in contrast with other extended objects like
the AdS black strings studied in Ref. \cite{PhysRevD.64.064010}
which can present Gregory-Laflamme gravitational instabilities
\cite{Gregory:1993vy} depending of the relation between the
longitudinal size of the horizon and the AdS radius.

\TABLE{
\begin{tabular}{ccccccc}
\hline 
\hline & \multicolumn{2}{c}{Tensorial} & 
\multicolumn{2}{c}{Vectorial}& 
\multicolumn{2}{c}{Scalar}
\\ \hline
$\;\;d\;\;$ & $\quad\;\;\mathfrak{w}_{\ss{R}}\quad\;\;$ &
$\quad\;\;\mathfrak{w}_{\ss{I}}\quad\;\;$ & $\quad\;\;
\mathfrak{w}_{\ss{R}}\quad\;\;$ &
$\quad\;\;\mathfrak{w}_{\ss{I}}\quad\;\;$& $\quad\;\;
\mathfrak{w}_{\ss{R}}\quad\;\;$ &
$\quad\;\;\mathfrak{w}_{\ss{I}}\quad\;\;$ \\
\hline 
4 & ---     & ---     & 1.84942 & 2.66385 & 1.84942 & 2.66385 \\
5 & 3.11945 & 2.74668 & 3.11945 & 2.74667 & 3.11945 & 2.74668\\
6 & 4.13591 & 2.69339 & 4.13591 & 2.69339 & 4.13591 & 2.69339 \\
7 & 5.00747 & 2.61247 & 5.00760 & 2.61266 & 5.00758 & 2.61249 \\
\hline\hline
\end{tabular}
\caption{The frequencies of the first non-hydrodynamic QNM for all
perturbation types, calculated with $\mathfrak{q}=0$. }
\label{tabq0}}

At this point we are able to confirm the agreement of our numerical results
with the analysis of the subsection \ref{subsecAsym}. We have seen in that
section that in the limit of small wavenumbers and large
frequencies the non-hydrodynamic QN frequencies are identical for all the
perturbation sectors. In order to see that we choose $\mathfrak{q}=0$ and
calculate the frequencies of the first non-hydrodynamic mode for each
perturbation sector and for $d=4$, $5$, $6$, $7$. The numerical data are
listed in Table \ref{tabq0}, from where we can see the very good
agreement between the analytical and the numerical results. Let us repeat
here that, since the Horowitz-Hubeny method presents convergence problems
for $d>6$, the results for $d=7$ in Table \ref{tabq0} were obtained
only by lowering the precision requirements. This explains the small
differences between the QN frequencies of each sector for $d=7$ in
Table \ref{tabq0}. Notice, however, that the results for tensorial
sector in $d=7$ spacetime dimensions are in agreement with the results
of Ref. \cite{Horowitz:1999jd}.

In Table \ref{check} some results obtained by both of
the numerical methods used in the present work, namely the
Horowitz-Hubeny method and the time domain evolution method, are shown
for comparison. We have chosen the fundamental mode ($n=0$) and the
number of dimensions where QNM frequencies were found through both of
the methods. It is seen that the two methods yield consistent and
satisfactory results for tensorial, vectorial and scalar type
perturbations.

\TABLE{
\begin{tabular}{ccccccc}
\hline 
\hline  &  &  & \multicolumn{2}{c}{Power series}
& \multicolumn{2}{c}{Time evolution}
\\ \hline
Type & $\;\; d \;\;$ & $\;\; \mathfrak{q} \;\;$  & $\quad
\mathfrak{w}_R\quad$ & $\quad \mathfrak{w}_I\quad$ &
$\quad \mathfrak{w}_R\quad$ &
$\quad \mathfrak{w}_I\quad$\\ \hline
& 5 & 10 & 11.0586 & 1.75039 & 11.0594 & 1.74773\\
Tensorial & 6 & 10  & 11.4325 & 1.70036 & 11.4313 & 1.69703\\
& 7 & 5 & 7.47908 &2.12521 & 7.48168 & 2.12835\\\hline
& 5 & 2  & 0 & 1.19612 & 0 & 1.19705\\
Vectorial & 6 & 2 & 0 & 0.872326 & 0 & 0.872788\\
& 7 & 2 & 0 & 0.700830 & 0 & 0.701345\\\hline
& 5 & 10 & 10.2220 & 0.536075 & 10.2713 & 0.677816\\
Scalar & 6 & 10 & 10.4304 & 0.728055 & 10.4320 & 0.730181\\
& 7 & 2 & 1.02097 & 0.520255 & 1.01691 & 0.533577
 \\ \hline\hline
\end{tabular}
\caption{\small{A comparison between some results obtained 
through the power series and the time evolution methods for 
$n=0$.}}\label{check}}

\section{QNM and the AdS/CFT correspondence}
\label{subsecAdSCFT}

\subsection{Thermalization timescale}

According to the AdS/CFT duality, perturbing a black hole in the AdS bulk
is equivalent to perturbing a CFT thermal state in the AdS spacetime 
boundary, and the time evolution of the black hole perturbation
describes the time evolution of fluctuations of the thermal state.
In particular, the characteristic damping time of a quasinormal mode,
$\tau=1/\omega_{\ss{I}}=(d-1)/(4\pi T\mathfrak{w}_{\ss{I}})$, is
related to the thermalization timescale of the
dual system, i.e., the characteristic time the perturbed thermal system
spends to return to thermal equilibrium. This timescale is dominated by the
quasinormal mode with lowest imaginary frequency.

In this subsection we study the decaying timescale of each
sector of perturbations. The most interesting case is the vectorial
sector, from which we begin the study.

\subsubsection{Vectorial sector}

Vectorial metric perturbations have two dominant quasinormal modes,
depending on the perturbation scale (wavelength): the hydrodynamic mode and
the first non-hydrodynamic mode. This happens because of a major difference
between the behavior of these modes. For the hydrodynamic mode one has
$\mathfrak{w}_{\ss{I}}\rightarrow \infty$ when $\mathfrak{q}$
increases to infinity. On the other hand, the non-hydrodynamic
mode behaves like $\mathfrak{w}_{\ss{I}} \rightarrow 0$ when $\mathfrak{q}$
increases. Then, we can see that from $\mathfrak{q}=0$ on the
thermalization timescale is dominated by the hydrodynamic mode, this is
true even when this timescale takes on the critical value, i.e., its lowest
value. Thereafter, the decaying timescale is dominated by the first
non-hydrodynamic mode.  Here we study this transition and the values of
$\mathfrak{q}$ and $\mathfrak{w}_{\ss{I}}$ where it occurs, as well as, the
value of the critical thermalization timescale in some dimensions. 

In the right panel of Fig.~\ref{figvet1vard} we have the hydrodynamic modes
together with the first non-hydrodynamic quasinormal modes for $d=4$, $5$,
..., $10$ dimensions. From the plots we can see that the hydrodynamic mode
dominates for small wavenumber and the non-hydrodynamic mode dominates for
intermediate wavenumber. For $d=6$ dimensions this behavior is better
observed in Fig.~\ref{figvet6d}, where one can see the transition from
non-oscillatory to oscillatory late-time decay. The explicit values for the
critical timescale in some dimensions are listed in Table \ref{tcritico}.
It is seen that the values for $\tau$ increase smoothly with
the number of dimensions $d$. The critical value for $\mathfrak{q}$ in
$d=5$ is consistent with the value found in Ref. \cite{Amado:2008ji}.

\TABLE{
\begin{tabular}{ccccccccccc}
\hline\hline
& \multicolumn{3}{c}{Vectorial} & \multicolumn{3}{c}
{Scalar}& \multicolumn{3}{c}
{Tensorial}\\
\cline{2-10}
$d$ & $\mathfrak{q}$ & $\mathfrak{w}_{\ss{I}}$ &
$\tau\,(\mbox{\small{\it{T}}}^{\,-1})$ & $\mathfrak{q}$ &
$\mathfrak{w}_{\ss{I}}$ &
$\tau\,(\mbox{\small{\it{T}}}^{\,-1})$ & $\mathfrak{q}$ &
$\mathfrak{w}_{\ss{I}}$ &
$\tau\,(\mbox{\small{\it{T}}}^{\,-1})$\\
\hline
4 & 1.935 & 2.29518 & 0.104015
& 3.213 & 0.414508 & 0.575942 & --- & --- & ---\\
5 & 2.622 & 2.40746 & 0.132218 & 3.360 & 0.913906 & 0.348296
& 0 & 2.69339 & 0.115889\\
6 & 3.100 & 2.38173 & 0.167058 & 4.081 & 1.66030 & 0.239648
& 0 & 2.74668 & 0.147727\\
7 & --- & --- & --- & --- & --- & ---
& 0 & 2.61200 & 0.182797\\
\hline\hline
\end{tabular}
\centering
\caption{The values of $\mathfrak{q}$ and $\mathfrak{w}_{\ss{I}}$ for
which $\tau$ assumes the minimum value for each sector of the
gravitational perturbations.}
\label{tcritico}}

\subsubsection{Scalar sector}

The relaxation timescale related to the scalar gravitational perturbations
in $d=4,\,5,\,6$ is dominated entirely by the hydrodynamic QNM.
However, the dispersion relation for the imaginary part of the hydrodynamic
frequency has a peak where $\mathfrak{w}_{\ss{I}}$ is maximum, so that in
this peak we have a critical (minimum) thermalization time (cf. Figs.
\ref{figscalarhydro} and \ref{figscal6d}). The critical values in this case
are shown in Table \ref{tcritico}. In this sector of gravitational
perturbations and for $d\le 6$, the values of $\tau$ decrease smoothly with
$d$, and we can see that this behavior is the opposite of that of the
vectorial and tensorial sectors.

\subsubsection{Tensorial sector}

Firstly we must remember that the tensorial modes arise only
in $d\geq 5$ spacetime dimensions and that the tensorial sector does not
present hydrodynamic modes, then the timescale is dominated by the first
non-hydrodynamic quasinormal mode whose behavior we can see in Figs.
\ref{figten1vard} and \ref{figten6d}. The absolute values of the
imaginary frequency decrease with $\mathfrak{q}$, so
that the critical timescale is in $\mathfrak{q}=0$. In this case
the values are listed in Table \ref{tcritico}.
In this sector the values for $\tau$ increase smoothly
with $d$, and this behavior is the same as that of the vectorial sector.

\subsection{Causality in the dual CFT plasma}

Recently some studies have used the wave-front velocity instead of the
group velocity in order to analyze the causality of signal propagation in
connection with the physics of the dual CFT plasma
\cite{Amado:2008ji,Natsuume:2007ty}. The wave-front velocity is considered
a reliable indicator if one wants to study how fast a signal can be
transmitted through a dispersive medium because it limits the speed of
propagation of a signal through the medium. The wave-front velocity is
defined as the velocity with which the onset of a signal travels
\cite{Amado:2008ji}: 
\begin{equation}
\label{frontvel} v_{\ss{F}}=\lim_{\mathfrak{q}\rightarrow\infty}
\frac{\mathfrak{w}}{\mathfrak{q}}. 
\end{equation} 
For $v_{\ss{F}}$ smaller than the speed of light, causality
is preserved (in this work, $c=1$). It also follows that the hydrodynamic
vectorial mode violates causality since from Eq.~\eqref{hidr} it has an
infinite limit for the wave-front velocity. However, it was shown here that
the vectorial hydrodynamic mode does not dominate in the large wavenumber
regime. Therefore one should analyze the first non-hydrodynamic QNM, which
is the dominant mode in the limit $\mathfrak{q}\rightarrow\infty$, and for
this mode it holds the general asymptotic formula
$\mathfrak{w}=\mathfrak{q}+ \alpha\mathfrak{q}^{-\beta}$, where
$\beta=(d-3)/(d+1)$ and $\alpha=\alpha_{\ss{R}}+i\alpha_{\ss{I}}$ is a
complex parameter depending on the number of dimensions $d$ and in the
overtone index $n$. Therewith, we found numerically that
$\lim_{\mathfrak{q}\rightarrow\infty}(\mathfrak{w}/\mathfrak{q})=1$. This
provides a proof of the causality of signal propagation in the dual CFT
plasma for any spacetime dimension.

\section{Final comments and conclusion}\label{secfinal}

In this work we have studied the complete quasinormal spectra of
gravitational perturbations of $d$-dimensional AdS
black branes.
Master equations for gravitational perturbations were derived for the
Kovtun-Starinets variables \cite{Kovtun:2005ev} in $d$ spacetime
dimensions, and, for comparison, the fundamental equations with the
Regge-Wheeler-Zerilli variables \cite{Kodama:2003jz} were also explored.
Among the relevant general results we can mention the proof
given in Section \ref{boundary} that for $d\ge 5$ dimensions, RWZ and KS
variables give the same QNM spectra. In this way we have unified two
different points of view for a consistent definition of gravitational
quasinormal modes: non-deformation of the boundary metric, associated to RWZ
variables, and the identification of QN frequencies with poles of
correlators, associated to KS variables.

Furthermore, the use of new gauge-invariant variables in
$d$-dimensional spacetimes allowed the calculation of the
hydrodynamic scalar and vectorial QNM of plane-symmetric AdS black
holes in $d$ spacetime dimensions. The expressions \eqref{hidr} and
\eqref{hidr2} are in complete agreement with the CFT predictions,
furnishing a non-trivial test to the AdS/CFT conjecture
\cite{Maldacena:1997re}. In addition, the results found here for
arbitrary $d$ reproduce exactly the results found in the literature
for $d=4$ and $d=7$ \cite{Herzog:2002fn,Herzog:2003ke,
Miranda:2008vb}, and for $d=5$
\cite{Policastro:2002se,Policastro:2002tn, Kovtun:2005ev}. Our
analytical and numerical calculations confirm the presence of
hydrodynamic (sound-wave) modes in any dimension.

The minimum value for the thermalization timescale was obtained for
$d=4,5,6$ and $7$. For the tensorial sector the thermalization time is
totally determined by the first non-hydrodynamic mode, while in the scalar
sector for $d=4,\,5,\,6$ it is dominated by the hydrodynamic
QNM alone. More interestingly, the vectorial sector for any $d$
and the scalar sector for $d\ge 7$ have two different dominant QNM,
the hydrodynamic mode for lower wavenumber values and the first
non-hydrodynamic mode for higher wavenumber values.

Even though the numerical methods presented some convergence problems, it
was possible to observe a general behavior of the QNM: the dispersion
relations for each sector are very similar for all the dimensions, except
for a few particularities specific to some specific mode and dimension.
In particular, we can mention the local ``knee'' appearing in the curve of
the real dispersion relation of the four-dimensional vectorial sector, and
the local maximum in the imaginary parts of frequency that appears for
$d>4$ in the scalar sector. Our analysis enables us to infer the absence of
tails in the time evolution profiles of gravitational perturbations of
non-extreme black branes \cite{Horowitz:1999jd}. However, the
presence of tails in extreme AdS black holes (and black branes)
spacetimes \cite{Ching:1995tj} is a problem which should be investigated in
detail in a future work.

It is also worth noticing that, as expected, our numerical results do not
show any instability of the black branes against tensor- and
vector-type perturbations. Moreover, for scalar-type
perturbations in $d\geq5$ our results suggest the stability of the
AdS black branes, what is an important result since the proof of such
a stability is still an open question \cite{Kodama:2003ck}.

Another important result is the confirmation that signal propagation in the
dual CFT plasma does not violate causality, independently of the number of
dimensions of the AdS spacetime. Although the wave-front velocity related
to the hydrodynamic vectorial mode grows with the wavenumber, the signal
propagation in the large wavenumber regime is dominated by the first
non-hydrodynamic vectorial mode, which obeys a relation of the form
$\mathfrak{w}=\mathfrak{q}+ \alpha\mathfrak{q}^{-\beta}$, with constant
$\alpha$ and non-negative constant $\beta$, resulting in the
wave-front velocity $v_{\ss{F}}=1$.

Finally, even though we have used two different numerical
methods, a power series and a time evolution methods, we cannot confirm
the existence of the highly real modes of Daghigh and Green
\cite{Daghigh:2009zz,Daghigh:2009fy}. The existence of such
QNM, and the reason why they appear in analytical studies but not in
numerical computations, are questions that should be addressed in
the future.

\section*{Acknowledgments}

This work is partially supported by Funda\c c\~ao para a Ci\^encia e
Tecnologia (FCT) - Portugal through project PTDC/FIS/64175/2006.
JM thanks Funda\c{c}\~ao Universidade Federal do ABC (UFABC) for a
grant. ASM, CM, and VTZ thank Conselho Nacional de Desenvolvimento
Cient\'\i fico e Tecnol\'ogico (CNPq) - Brazil for grants.

\end{document}